\documentclass{aa}
\usepackage{graphicx}
\usepackage{txfonts}
\usepackage{longtable}
\usepackage{afterpage}
\usepackage{rotating}
\begin{document}%
\title{Integral field spectroscopy based H$\alpha$ sizes of  local luminous and ultraluminous infrared galaxies}

\subtitle{A direct comparison with high-z massive star forming galaxies}

\author{
S. Arribas\inst{1}, L. Colina\inst{1}, A. Alonso-Herrero\inst{1,2}, F.F. Rosales-Ortega\inst{1,3}, A. Monreal-Ibero\inst{4}, M. Garc{\'{\i}}a-Mar{\'{\i}}n\inst{5}, S. Garc{\'{\i}}a-Burillo\inst{6}, J. Rodr\'iguez-Zaur{\'{\i}}n\inst{7}}
 \offprints{arribas@cab.inta-csic.es}

\institute{
Centro de Astrobiolog{\'{\i}}a,  Departamento de Astrof\'isica, CSIC-INTA, Ctra. de Ajalvir km. 4, 28850 -Torrej\'on de Ardoz, Madrid, Spain. \\\email{arribas@cab.inta-csic.es}          
\and
Instituto de F\'isica de Cantabria, CSIC-UC, Avenida de los Castros S/N, 39005-Santander, Spain
\and
Departamento de F{\'i}sica Te{\'o}rica, Universidad Aut\'onoma de Madrid, 28049 Madrid, Spain.
\and
Instituto de Astrof\'isica de Andaluc\'ia. CSIC. Glorieta de la Astronom\'ia, s/n, 18008, Granada, Spain.
\and
Physikalisches Institut, Universit{\"a}t zu K{\"o}ln, Z{\"u}lpicher Strasse 77, 50937 K{\"o}ln, Germany
\and
Observatorio Astron\'omico Nacional (OAN)-Observatorio de Madrid, Alfonso XII, 3, 28014-Madrid, Spain
\and
Space Telescope Science Institute, 3700 San Martin Drive, MD 21218, Baltimore, USA
}

\date{}

 
  \abstract
   {}
   {We study the analogy between local luminous and ultraluminous infrared galaxies (U/LIRGs) and high-z massive star forming galaxies (SFGs) by comparing their basic H$\alpha$ structural characteristics, such as size and luminosity surface density, in an homogeneous way (i.e. same tracer, size definition, and similar physical scales). }
   {We use integral field spectroscopy (IFS)  based H$\alpha$ emission maps for a representative sample of 54 local U/LIRGs (66 galaxies) observed with INTEGRAL/WHT and VIMOS/VLT.  From this initial sample, we select 26 objects with similar H$\alpha$ luminosities (L(H$\alpha$)) to those of massive (i.e. M$_\star$ $\sim$ 10$^{10}$ M$_\odot$ or larger) SFGs at z $\sim$ 2, and observed on similar physical scales.  We then directly compare the sizes, and luminosity (and SFR) surface densities of these local and high-z samples.}
   {The size of the H$\alpha$ emitting region in the local U/LIRGs that we study has a wide range of values, with r$_{1/2}$(H$\alpha$) from 0.2 kpc  to  7 kpc. However, about two-thirds of local U/LIRGs with L$_{ir} >$ 10$^{11.4}$ L$_\odot$ have compact H$\alpha$ emission (i.e. r$_{1/2}$ $<$ 2 kpc). The comparison sample of local U/LIRGs also contains a larger fraction (59$\%$) of objects with compact H$\alpha$ emission than the high-z sample (25$\%$).  This gives further support to the idea that for this luminosity range the size of the star forming region is a distinctive factor when comparing local and distant galaxies of similar SFRs. However,  when using H$\alpha$  as a tracer for both local and high-z samples, the differences are smaller than those reported using a variety of other tracers. In the L(H$\alpha$)  -  L(H$\alpha$) surface density ($\Sigma_{H\alpha}$) plane, most of the local U/LIRGs and high-z SF galaxies follow the same trend (i.e. higher luminosity for higher surface density) and cover a similar range, except for about 20-40 $\%$ of  local U/LIRGs, which have a higher $\Sigma_{H\alpha}$ by a factor of about 10.  This is considerably smaller than the factors of 1000 or more reported in similar planes  (i.e. L(TIR) versus  $\Sigma_{TIR}$).  Despite of the higher fraction of galaxies with compact H$\alpha$ emission, a sizable group (about one-third) of local U/LIRGs are large (i.e.  r$_{1/2}$ $>$ 2 kpc).  These are systems that show evidence of pre-coalescence merger activity and are indistinguishable from the massive high-z SFGs galaxies in terms of their H$\alpha$ sizes, and luminosity and SFR surface densities.}
   {} 
 \maketitle
%

\section {Introduction}

Luminous and ultraluminous infrared galaxies (LIRGs: $L_{ir} \equiv L[8-1000 \mu m]= 10^{11-12}L_{\odot}$; ULIRGs: $L_{ir} > 10^{12}L_{\odot}$) are believed to have an important role in our understanding of galaxy evolution. They are systems of intense star formation (SF), whose spectral energy distributions (SEDs) are dominated by dust thermal emission arising from the reprocessing of UV photons produced by young massive stars and/or active galactic nucleus (AGN) heating. The fraction of these systems with disturbed morphologies increases with luminosity, with most ULIRGs showing evidence of recent or on-going merger events. By comparison, LIRGs seem to be a more heterogeneous group with in many cases properties similar to those of isolated star-forming spirals, especially at low luminosities (i.e.  $L_{ir} \sim 10^{11}L_{\odot}$).  Although U/LIRGs are rare locally, studies with the {\it Spitzer Space Telescope} have shown that they are much more numerous at high-z and account for an increasingly larger fraction of the total star formation density fraction (e.g. more than half of the total SF density at z=2, P\'erez-Gonz\'alez et al, 2005; although see Rodighiero et al. 2011). The so-called sub-millimeter galaxies (SMGs; e.g. Smail et al. 1997) are commonly viewed as a more luminous counterpart of local U/LIRGs. 

However, although many high-z galaxies meet the above (luminosity) definition, the analogy between local and distant LIRGs and ULIRGs is under discussion. Several authors have reported that the SEDs of high-z U/LIRGs are similar to lower-luminosity local systems suggesting that they are not high-z analogs but instead scaled-up versions of lower-luminosity local U/LIRGs (e.g. Pope et al. 2006, Papovich et al. 2007, Takagi et al. 2010, Muzzin et al. 2010).

It has been suggested that the observed differences between local and distant U/LIRGs (of similar luminosity) could be due to a difference in metallicity and/or physical size (e.g. Rigby et al. 2008; Farrah et al. 2008).  On the one hand, Engelbracht et al. (2008) find that the differences between the SEDs of local and z $\sim$ 2 ULIRGs are qualitatively consistent with a difference in metallicity of a factor 1.5-2. On the other hand,  several authors find that local U/LIRGs have a more compact structure than high-z populations  (e.g. Iono et al. 2009,  Rujopakarn et al. 2011, and references therein).  Elbaz et al. (2011) also linked the differences in SEDs to the compactness of the star-forming region. 

Using a variety of tracers, Rujopakarn et al (2011) report that local U/LIRGs have smaller SF region sizes by up to factors  of 50 (i.e. more than 3 orders of magnitude in luminosity surface density) than other local and high-z SFGs.  Therefore, they suggest that local ULIRGs and LIRGs belong to a rare population driven by a unique process. These results have brought a lot of attention to the determination of U/LIRG  sizes (especially for local samples).  

Despite size being a fundamental  property, its determination is often uncertain as it is affected by a number of observational and instrumental factors (e.g. reddening, resolution, etc). In addition, different spectral features trace different galaxy components and physical mechanisms. This complicates the comparison between local and distant objects, which are often selected and observed at different (rest-frame) wavelengths. Differences in methodology also add uncertainty to this comparison.  

This paper compares some structural properties of the H$\alpha$ emitting region in local U/LIRGs, such as sizes and luminosity surface densities,  with those of high-z populations.  We use integral field spectroscopy (IFS) based  H$\alpha$ emission maps obtained from our INTEGRAL/WHT and VIMOS/VLT observations to derive these properties for a representative sample of local U/LIRGs (e.g. Colina et al., 2005; Garc{\'{\i}}a-Mar{\'{\i}}n et al, 2009a; Rodr{\'{\i}}guez-Zaur{\'{\i}}n et al, 2011, and references therein).  We compare our results with those for high-redshift massive SFGs samples observed with near-infrared IFS on similar linear scales, using the same tracer (e.g. F{\"o}rster-Schreiber et al, 2009, 2011; Wright et al., 2009; Law et al., 2009; Wisniosky et al. 2011).  Therefore, we follow similar methods for the local and distant samples to minimize the uncertainties in the relative comparison. As we base our size measurements on reddened H$\alpha$ maps, they refer to the extension of the "unobscured"  H$\alpha$-emitting region.  

The paper is structured as follows. In Section 2, we briefly describe the IFS sample and the data used for the analysis.  In Section 3, we first derive SF region sizes from H$\alpha$ for the whole sample of U/LIRGs observed with INTEGRAL/WHT and VIMOS/VLT. From this group, we then select the subsample that allows us to make a homogeneous comparison with the high-z samples.  The sizes and luminosity surface densities of distant and local U/LIRGs samples are confronted in Sec. 4.  Finally our main conclusions are summarized in Section 5.  Throughout the paper, we consider H$_{0}$ = 70
kms$^{-1}$Mpc$^{-1}$, $\Omega_{\rm \Lambda}$ = 0.7, $\Omega_{\rm M}$ = 0.3.

\section {The sample and observations} 

The sample for the present study is formed by local LIRGs and ULIRGs for which we have obtained optical IFS data with the VIMOS-IFU (LeF$\grave{e}$vre et al. 2003) and the INTEGRAL (Arribas et al. 1998) instruments.  The objects were selected to form a representative sample of U/LIRGs covering all types of morphologies (isolated spirals, interacting pairs, mergers), nuclear excitation (HII, Seyfert, and LINER), and sampling the LIRG and ULIRG infrared luminosity range.  The sample is incomplete in either flux or distance, since complete samples covering the U/LIRG  luminosity range would be so numerous that their IFS observations would require a prohibitively large amount of time.    

The INTEGRAL sub-sample is that presented in  Garc{\'{\i}}a-Mar{\'{\i}}n et al. (2009a).  These are northern objects selected  from the IR-bright samples of  Sanders et al. (1988), Melnick \& Mirabel (1990), Leech et al. (1994), Kim et al. (1995), Lawrence et al. (1999), and Clements et al. (1996).
For the present analysis we exclude IRAS F09427+1929 and IRAS F13469+5833, for which the H$\alpha$ data have low S/N, Mrk231 which is severely contaminated by an AGN, and IRAS F13342+3932, which has another AGN and it is observed near the edge of the FoV and therefore a reliable size determination cannot be made. In general, the INTEGRAL observations were carried out with bundle SB2 (i.e. optical fibers of 0.9 arcsec in diameter and FoV of 12.3 x 16 arcsec$^2$), except for a few cases (see Table 2). The spectral resolution is about 6 \AA. Details about the observations, reductions, and calibrations of these data can be found in Garc{\'{\i}}a-Mar{\'{\i}}n et al (2009a, hereafter GM09). 

The VIMOS sample is presented in Rodr{\'{\i}}guez-Zaur{\'{\i}}n et al (2011) and is drawn from the IRAS Revised Bright Galaxy Sample (Sanders et al. 2003), the IRAS 1 Jy sample of ULIRGs (Kim et al. 1998), and the HST/WFPC2 snapshot sample of bright ULIRGs (ID 6346 PI: K.Borne). These are mainly southern objects. For the present analysis, we exclude F08424-3130 and F12596-1529 because they are not well-covered by the instrument FoV. We used the high resolution mode with the HR-orange grating, which provides a spectral resolution of about 2 \AA.  After combining four dither pointings, the total FoV is about 30 x 30 arcsec$^2$, with a spaxel scale of 0.67 arcsec (square). Details about the observations, data reduction, and calibration can be found in Monreal-Ibero et al. (2010) and Rodr{\'{\i}}guez-Zaur{\'{\i}}n et al. (2011, hereafter RZ11). 

In summary, the whole sample consists of 54 systems (66 individual galaxies)\footnote{ When clearly distinct, individual galaxies in multiple systems are treated independently.}, 32 LIRGs, and 22 ULIRGs.  Their mean (median) distance is 226 (137)  Mpc,  covering a range from 40.4 Mpc ( z= 0.0093) to 898  Mpc (z=  0.185).  Their infrared luminosity spreads over the range  $10^{10.8}L_{\odot} < L_{ir} < 10^{12.6}L_{\odot}$. It also includes all types of nuclear excitations and interaction phases, and therefore it should be representative of the general properties of local U/LIRGs.

\section{H$\alpha$ emitting region extension in local U/LIRGs}

 
Leaving aside AGN and shock effects, the H$\alpha$ emission traces the (moderately obscured) gaseous regions ionized by young massive stars. Obtaining H$\alpha$ emission maps is generally costly as it requires the use of some kind of 3D observational technique. For the present study, we obtain the H$\alpha$ maps from the IFS data cubes (see Sec. 2) after fitting the H$\alpha$-[NII] complex in the individual spectra associated with each spaxel (GM09, RZ11). From these maps, we derive half-light radii (i.e. r$_{1/2}$).

A method commonly used to infer half-light radii is based on fitting the observed flux distribution to a galaxy model assuming some standard surface brightness profiles (e.g. GALFIT, Peng et al. 2010). This method is accurate as long as the model is a good representation of the actual galaxy flux distribution. This is usually the case when studying the starlight of the stellar population traced by the rest-frame optical and near-infrared continuum in  {\it normal} galaxies. However, the star forming regions traced by their H$\alpha$ emission have usually irregular, clumpy, ring-shaped,  or other peculiar morphologies that cannot be adequately modeled with standard profiles. This is especially true for interacting and merger systems such as U/LIRGs observed at relatively high spatial resolutions (i.e. sub-kpc scales). In these cases, the half-light radii can be obtained from the curve-of-growth (CoG) of the flux in increasingly larger apertures. We follow this approach for the local sample  (Table 2), as most previously studied massive high-z galaxies have determinations based on this method. In addition, this method depends less on angular resolution effects than others (F{\"o}rster Schreiber et al, 2009).  However, it is important to realize that the r$_{1/2}$ (CoG) of the H$\alpha$ emitting region is not necessarily a measure of the {\it absolute} size of the starburst (if such a concept exists for a clumpy flux-uneven structure). These measurements should only be considered within the constraints imposed by the tracer, angular resolution, and radius definition (e.g. CoG) used. These H$\alpha$ determinations for local samples are of relevance because similar measurements are being obtained for high-z samples, hence a  direct comparison is possible.

In Table 2, we also include size estimates obtained using an alternative method (which we refer to as the A/2 method) that computes r$_{1/2}$=$\sqrt {A/\pi}$, where A is the angular extent of the minimum number of pixels (or spaxels) encompassing half of the flux. The results of this method do not depend on the location of individual emitting regions across the FoV, but on their actual extent. The method does not require knowledge of the  galaxy center, which in some cases is offset between the different galaxy components (i.e.  H$\alpha$ and the continuum emission; e.g.  Garc{\'{\i}}a-Mar{\'{\i}}n et al. 2009b). This method has been previously used in high-z samples (e.g. Erb et al. 2004), though the CoG is more commonly applied.  The CoG and A/2 methods provide similar results for a smooth radially decaying flux emission, but differ for highly structured objects. The detailed comparison among different tracers and methods for estimating sizes in SFGs will be made somewhere else. Here we derive H$\alpha$ sizes based on the CoG method in order to homogenize the comparison with high-z studies. 

The sizes obtained directly from the H$\alpha$ images were transformed into intrinsic (i.e. deconvolved) sizes by subtracting in quadrature the PSF (Table 2),  as done for high-z samples.  In Figure 1, we present these results as a function of the infrared luminosity. We distinguish between interacting and pre- coalescence merger systems (triangles), post-coalescence mergers (squares), and isolated disk galaxies (circles).  Five objects have half-light radii equal to or smaller than the PSF, and are represented as upper limits.  Measured sizes smaller than the PSF can be obtained as a consequence of seeing fluctuations. For a relatively large fraction of LIRGs only a lower limit to the size could be obtained.  These are objects that, after visual inspection of their H$\alpha$ maps, have clear evidence that a fraction of the emission is lying  outside the IFU frame.  

The  sample shows a wide range of sizes, from r$_{1/2} \sim$ 0.2  up to 7 kpc (see Tables 1 and 2). To analyze the size distribution, we restrict the sample to objects with log (L$_{ir}$/L$_\odot$) $>$ 11.4 because the problems due to the limited  FoV are significantly smaller.  This group contains a large fraction of objects with small sizes:  39 $\%$  (16/41) have r $_{1/2}$ $<$ 1 kpc, and 68 $\%$ (28/41) r$_{1/2}$ $<$ 2 kpc. However, a sizable fraction (13/41) of objects have large H$\alpha$ emitting regions (i.e. r$_{1/2} >$ 2 kpc).  Most of these large objects are ULIRGs  in a  pre-coalescense  merger phase, with nuclear separations ranging from 1.5 kpc to 14 kpc (GM09a).  The only  {\it exception} is the ULIRG IRAS 11087+5351, which has a nuclear separation of 1.5 kpc, at the border considered by GM09a to distinguish pre- and post-coalescence systems.  The characterization of the merger phase in these ULIRGs is secure in most of the cases thanks to HST imaging, which also reveals large (30-50 kpc) envelopes associated with the old stellar component and/or with prominent tidal structures (GM09a; see also Sec. 4.1). 

For many objects with log (L$_{ir}$/L$_\odot$) $<$ 11.4 we could only infer lower limits to the sizes of the H$\alpha$ emitting region because of the limited FoV of the IFS instruments.This luminosity range includes objects with a wide range of morphologies (e.g. Arribas et al. 2004), and extended H$\alpha$ emission is not necessarily associated with mergers, but with isolated disks as well. We note that, on the basis of  NICMOS imaging of a distance limited sample of 30 LIRGs,  Alonso-Herrero et al. (2006) found that about half of the sample have compact (1-2 kpc) Pa$\alpha$ emission with a high surface brightness, while for the remaining half the emission extends over scales of 3-7 kpc and larger.

\begin{table*}
\caption{H$\alpha$ half-light radius  and  luminosity and SFR surface density for different local U/LIRGs and high-z samples.}\label{table:tabla1}
\begin{tabular}{l c c c c c c c c c c c  }
\hline \hline
Sample            & n &  \multicolumn{3}{c} {r$_{1/2}$(H$\alpha$)}                &   \multicolumn{3}{c} {log$\Sigma_{H\alpha}$}&  \multicolumn{3}{c} {$\Sigma_{SFR}$} & Comments \\               
                          &    &  \multicolumn{3}{c} {(kpc)}                                             &   \multicolumn{3}{c} {(erg s$^{-1}$kpc$^{-2}$)}&  \multicolumn{3}{c} {(M$_\odot$yr$^{-1}$kpc$^{-2}$)}  \\   \cline{3-4} \cline{6-7}  \cline{9-10}                   
                          &  & Median &  Range & & Median & Range & & Median & Range & &  \\  \cline{3-4} \cline{6-7} \cline{9-10} 
Local U/LIRGs  & & & & & & & & & & & i\\  
\hline
\hline 
log(L$_{ir}$/L$_\odot$)$>$11.4 (all) & 41 &  1.1 & 0.2 - 7.1 & & 41.6 & 40.5-43.1 & & 3.1  & 0.2-91.& &  ii \\
log(L$_{ir}$/L$_\odot$)$>$11.4 (no AGN)) & 33 & 1.0 & 0.2 - 7.1 & & 41.4& 40.5-43.1 & & 2.2 & 0.2-91.& & iii\\
log(L$_{ir}$/L$_\odot$)$<$11.4 (all) & 20   & $>$1.7 & 0.2 - $>$3.7 & & $<$40.6& $<$39.6-42.4& & $<$0.3 & $<$.03-18. & & iv\\
\hline
Local U/LIRGs subsample & & & & & & & & & & & v\\
for comparison with high-z & & & & & & & & & & &   \\
\hline 
\hline
All & 26  & 1.9 & 0.3 - 7.1 & & 41.6& 40.5-43.1& & 3.1 & 0.2-91.&& \\
No AGNs & 20 & 1.6 & $<$0.5 - 7.1 & & 41.7& 40.5-43.1& &3.9 & 0.2-91.&& iii \\
No pre-coalescence  & 10 & 1.0 & 0.3 - 1.9 & & 42.2& 41.1-43.1& & 14. & 1.1-91.& & vi\\
Only pre-coalescence & 16 & 2.9 & 0.4 - 7.1 & & 41.2& 40.5-42.6&& 1.1 & 0.2-34.& & vii\\
\hline
High-z SGFs   & & && & & & & & & & viii\\
\hline
\hline
All & 81  & 2.8 & 0.6 - 7.5 & & 41.5& 40.2-42.8& & 2.7 & 0.1-50. &&  \\
IR selected & 34 & 3.4 & 1.3 - 7.5 & & 41.5& 40.2-42.6& & 2.7 & 0.1-34. &&  ix \\
\hline
\hline
\end{tabular}
\tablefoot{ 
Main columns are: (1) Sample: For the characteristics of the different samples considered see comments in column (6) and main text; (2) n: number of objects in each sample;  (3) r$_{1/2}$(H$\alpha$): intrinsic (deconvolved) H$\alpha$ half-light radius; (4)  logarithm of  the dereddened  H$\alpha$ luminosity surface density within the half-light radius, r$_{1/2}$(H$\alpha$). Note that  $\Sigma_{H\alpha}$ = L(H$\alpha$)/(2$\times$$\pi$r$_{1/2}^2$), where L(H$\alpha$) is the total H$\alpha$ luminosity and the factor 2 takes into account that within the half-light radius only half of the flux is included; (5) Star formation rate surface density within r$_{1/2}$(H$\alpha$), obtained from $\Sigma_{H\alpha}$ following Kennicutt (1998); (6) Comments where:  (i) Sample observed via IFS with INTEGRAL and VIMOS-IFUs. Mrk 273 is not included in the luminosity calculations, as no IFS calibrated data exists (GM09b). For Arp 220, the H$\alpha$ luminosity was obtained from Colina et al. (2004), while for the rest of the sample they come from GM09b and RZ11 (see text).  This sample is divided at  log(L$_{ir}$/L$_\odot$)$=$11.4 because r$_{1/2}$ values for the subsample above this limit are virtually unaffected by FoV, while for a large fraction of objects below this luminosity  only lower limits to their size could be obtained (see text, and Fig. 1).  (ii) Galaxies C and S of  IRAS 06259-4708 are not included  as it is uncertain whether they belong to this group as no reliable individual values for their infrared luminosity exits.  In any case, the reported statistical values do not change by their inclusion. (iii) Objects with evidence of hosting an AGN have been removed from this sample. (iv) For a large fraction of objects in this luminosity range, the r$_{1/2}$ determinations are significantly affected by the small FoV of the IFS instruments used and, therefore, these values should be only considered as lower limits. (v) Local U/LIRGs with (dereddened) L(H$\alpha$) $>$ 10$^{42}$ ergs$^{-1}$ and observed on similar linear scales as the selected high-z samples. (vi) This set excludes objects classified as interacting and pre-coalescence systems. IRAS F11087+5355, which is at the border of the definition of a pre-coalescence system  (see GM09a) is also excluded. (vii) Only interacting and pre-coalescence systems (IRAS F11087+5355 is also included). (viii) High-z SFGs observed in H$\alpha$ with IFS systems with high angular resolutions. They include the works of: F{\"o}rster Schreiber et al. (2009), Law et al. (2009), Wright et al. (2009), and  Wisnioski et al. (2011). (ix) This sample includes only the IR-selected sources of the SINS sample of F{\"o}rster Schreiber et al. (2009).
}
\end{table*}

  \begin{figure}[h]
 \centering  

 \includegraphics[width=8cm]{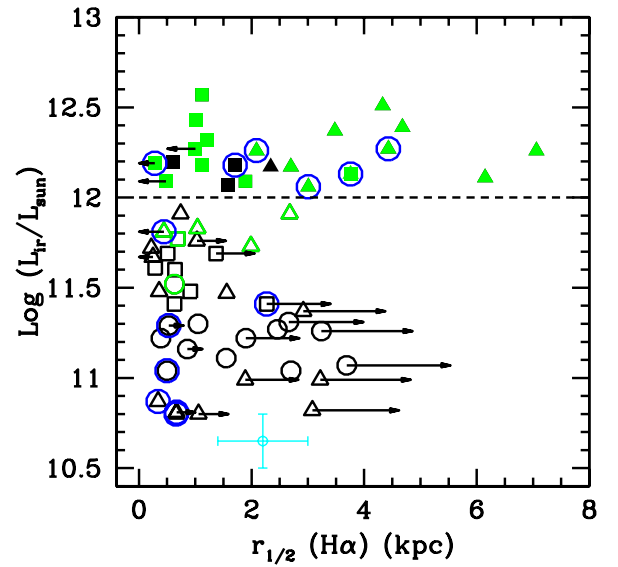}
 \caption{Intrinsic H$\alpha$ half-light radii (CoG, method)  of local U/LIRGs as a function of the infrared luminosity. The symbols indicate different  morphological classes: triangles are interacting and pre-coalescence systems, squares are post- coalescense mergers, and circles are isolated disk galaxies. Green symbols are objects included in the subsample for comparison with high-z.  The individual galaxies of the system IRAS 06259-4708 (C and S), which also belong to this subsample, are not included in the plot since no values for their L$_{ir}$ exist. However, considering the total luminosity of the system (log(L$_{ir}$/L$_\odot$)=11.91) and their H$\alpha$ luminosities, they are likely LIRGs.  Open symbols are used for LIRGs, and solid ones for ULIRGs. Values encircled in blue indicate targets with evidence of hosting a (weak) AGN according to their optical spectra (see Table 2). For {\it unresolved} systems, upper limits are represented. For some  low luminosity objects, the values represented are lower limits owing to the limited FoV. This is illustrated by the arrows, which indicate a factor of 1.5 increase in size. The blue cross in the lower-right corner indicates the typical errors of an object with r$_{1/2}$ $\sim$ 2 kpc (for detailed individual errors see Table 2). See text. }
\
 \label{Fig2}
 \end{figure}

 \subsection{Reddening effects} 
 
A radial variation in the extinction modifies the flux distribution and therefore affects the radius determinations. The U/LIRGs have large amounts of dust in their nuclear regions (e.g. GM09b), and therefore extinction-corrected half-light radii should be smaller than uncorrected ones.  However, detailed reddening corrections of the radius determinations are beyond our current possibilities. Aside from the Balmer decrement providing only a partial estimate of the actual extinction (e.g. Alonso-Herrero et al. 2006),  the  2D reddening structure is mostly unknown. In general, H$\beta$, and therefore the Balmer decrement, can only be obtained for a few objects observed with  INTEGRAL (GM09a), and is restricted to the innermost regions where H$\beta$ is detected with sufficient S/N. Hence, this index cannot be used to map the extinction over a sufficiently large area to correct individually the radii. However, these IFS data provide reliable individual nuclear extinction values, as well as some general constraints on the extranuclear region.  In particular, GM09b found that extranuclear extinction for ULIRGs is patchy, with a large scatter among objects and, in general, undergoes a radial decay. 

To obtain a coarse estimate of the importance of the extinction effects on our determinations, we correct our images with a simple model of extinction consisting of a linear decay from the nuclear value of Av (inferred from the Balmer decrement) up to Av=0 in the outermost regions of the extended H$\alpha$ emission. This implies gradients in the range of 0.15 - 1 visual magnitudes of extinction per kpc. This model has two main limitations. First, it does not capture the patchy nature of the extinction. Second, it only considers the extinction inferred from the Balmer decrement, which as mentioned above gives only a lower limit to the true obscuration. In principle, the limitations of using the Balmer decrement should be greater for objects with high extinction in their nuclei. However, this model can provide us with a reference for evaluating the importance of extinction effects. When applying this correction to the sample \footnote{ Except for compact objects (i.e. observed half-light radius $\le$ 3 spaxels) to avoid uncertainties associated with the deconvolution.}, we found that r$_{1/2}$(Av-corrected) / r$_{1/2}$ = 0.72 $\pm$ 0.15 (also see Fig. 2 for the distribution of values), and therefore a mean reduction in size of about 25-30 percent.  We attempted to analyze a possible correlation between these correction factors  and L$_{ir}$, but the low statistics and the large scatter among the objects prevent us from reaching any firm conclusion. 

In this context, it is also relevant to mention that Alonso-Herrero et al. (2009) found that ground-based IFS H$\alpha$ and NICMOS Pa$\alpha$ emissions of LIRGs have similar morphologies, suggesting that the extinction effects on H$\alpha$ are not severe, except in the very nuclear regions.

 \begin{figure}[h]
 \centering  

 \includegraphics[width=8cm]{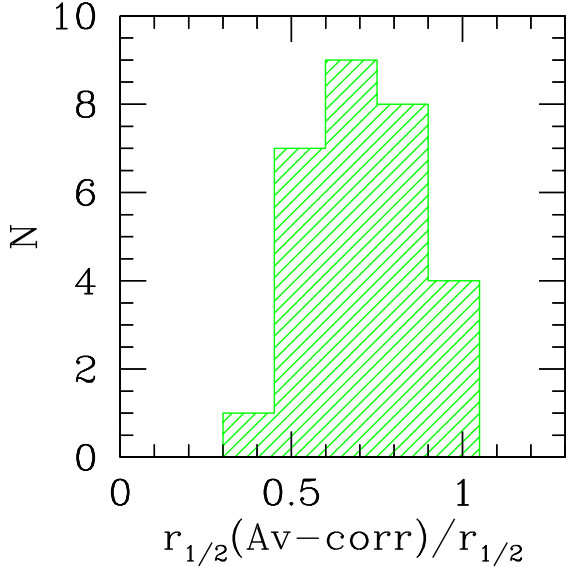}
 \caption{ Distribution of the ratio between the half-light radii obtained after correcting from extinction with a simple model (see text), and the ones derived directly without correction.}
\
 \label{Fig2b}
 \end{figure}
            
As for the continuum emission,  Veilleux et al. (2002) obtained ground-based optical  {\it R} and near-infrared  {\it K} imaging for most of the 118 ULIRGs from the IRAS 1 Jy sample, and inferred mean half-light radii of 4.80 $\pm$ 1.37 and 3.48 $\pm$ 1.39 kpc, respectively. This difference (38\%) cannot be attributed totally to the difference in extinction between the two bands, since looking at the original images (Kim et al. 2002) it is clear that the K-band images are considerably shallower, which make their size determinations relatively smaller.  

Summarizing, though a detailed size correction for reddening is impossible, our simple model as well as  other empirical results (Alonso-Herrero et al. 2006 and Veilleux et al. 2002) do not  suggest that the (observed) reddening has a severe impact on the radius determinations.  For the remainder part of the paper, we use uncorrected size measurements, and therefore they refer to the extension of the unobscured H$\alpha$ emitting region.  

\subsection{ H$\alpha$ versus MIR emission extension}

A detailed comparison of our  H$\alpha$ sizes with previous MIR measurements is difficult because of the different methodology, angular resolution, and size definition used by the different works, and is beyond the goals of this paper. However, the general conclusions show relatively good agreement. On the one hand, D\'iaz-Santos et al. (2010), who used long-slit  IRS spectra to estimate the fraction of  extended emission, showed that the MIR continuum (i.e. 13.2 $\mu$m) in LIRGs originates on scales of up to 10 kpc, with a mean size of the cores of 2.6 kpc.  These figures are consistent with our findings, especially taking into account the different size definition and angular resolution.  They also find that for ULIRGs the MIR emission is more compact.  Our data do not allow us to study in detail the dependence of size on luminosity, because for many of the low luminosity objects we could only infer lower limits to their size. However, the comparison of the median values of our high and low luminosity bins (see Table 1), suggests a similar behavior to that of D\'iaz-Santos et al. (2010).  Finally, they also found that the compactness strongly increases in objects classified as mergers in their final stage of interaction. This is also clearly observed in our sample when comparing pre- and post- coalescence systems (Fig.1).  D\'iaz-Santos et al. (2011) found that the [NeII]12.8 $\mu$m emission is as compact as the continuum dust emission, though the polycyclic aromatic hydrocarbon (PAH) emission is more extended.

In addition, there is evidence that the H$\alpha$,  Pa$\alpha$, and MIR emissions in LIRGs are strongly correlated with each other. On the one hand, Alonso-Herrero et al. (2006b) and D\'iaz-Santos et al. (2008) showed that the overall morphologies of the MIR 8$\mu$m emission (produced by thermal continuum from hot dust and by a PAH feature) and the Pa$\alpha$ emission line of LIRGs are similar. Moreover, Alonso-Herrero et al. ( 2009) found similar IFS H$\alpha$ and NICMOS Pa$\alpha$ morphologies, both tracing the nuclear emission as well as the emission from bright high surface-brightness HII regions. In addition, Alonso-Herrero et al. (2006b) showed that the fraction of total emission contained in the relatively small NICMOS FoV (19$''$$\times$19$''$) is similar for  H$\alpha$ and 24$\mu$m for three LIRGs (see their table 7), suggesting further that a large difference among the global extension derived from H$\alpha$ and the MIR emission is not expected.  

 \subsection{AGN effects} 
 
The presence of a bright AGN could affect the size determination, as the extra flux in the nuclear regions associated to the AGN would reduce the derived  half-light radius. From the original INTEGRAL and VIMOS samples, several objects exhibiting a strong influence by the AGN on their optical spectra were excluded (e.g. Mrk231,  IRAS F13342+3932). However, we retain other objects with hints of a weak or modest AGN contamination (see Table 2),  for which the radius measurements should not in principle be severely affected. We note that removing all objects with evidence of an AGN may also introduce some systematic biases in U/LIRG samples because the presence of an AGN correlates with other object properties (e.g. Veilleux et al. 1999).  However, we mark these objects in the figures and discuss the possible influence of the AGN on specific results. This approach is similar to that of other studies of high-z samples (e.g. FS09), which are known to be contaminated by AGNs (see also Shapiro et al, 2009).  
 
Fig. 1 shows that those galaxies with evidence of hosting an AGN do not have a significantly distinct behavior from the rest of the sample. As can also be seen in Table 1, this is particularly true for the subsample of galaxies with log(L$_{ir}$/L$_\odot$) $>$ 11.4, and for that used for high-z comparison  (see Sec. 4.1). However, for the low luminosity bin (i.e. L$_{ir}$/L$_\odot$) $<$ 11.4), objects with an AGN seem to be on average smaller than those without traces of activity. This result should be interprted with caution as it may be a consequence of the small number statistics.          

In this context, we consider the results of Alonso-Herrero et al (2012) who, based on Spitzer data, found that only 8$\%$ of local LIRGs have a significant AGN bolometric contribution L$_{bol}[AGN]$/L$_{ir}$ $>$ 0.25. For ULIRGs, the effects are expected to be somewhat larger as a consequence of the well-known trend of increasing AGN significance with bolometric luminosity (e.g. Veilleux et al. 1999, Nardini et al. 2010). 

 \section{Direct comparison with high-z samples} 
 
\subsection {IFS based H$\alpha$ emission size of local and high-z SFGs samples} 

One should ideally use not only the same tracer, but also data of similar linear resolutions when comparing the sizes of SF regions of different galaxy samples.  This is in general difficult  when comparing local and high-z samples owing to the different observed-frame wavelengths of the tracer, and the generally different linear resolutions on target.

Major efforts have been made to characterize high-z SFG galaxies by different groups, and a number of samples  of up to z $\sim$ 2.5  have been observed in H$\alpha$  (e.g. Erb et al. 2006a,b, Genzel et al. 2008, Wright et al. 2009, Law et al. 2009, F{\"o}rster-Schreiber et al. 2009, 2011, Epinat et al. 2009, Jones et al. 2010, Wisnioski et al. 2011, Nelson et al. 2012, Sobral et al. 2012 and references therein). Some of these studies are based on near-IR IFS with high angular resolution (i.e.  either AO-assisted or under good seeing conditions), providing sub- to few kpc linear resolutions.   Our seeing-limited optical IFS observations probe the H$\alpha$ emission of local U/LIRGs at linear resolutions ranging approximately from 0.2 kpc to 4 kpc.  Therefore, there is an overlapping  fraction of local and distant galaxies for which  H$\alpha$ sizes can be compared in a rather homogeneous way, with of similar linear resolutions.

\begin{figure*}[]
\centering  
  \includegraphics[width=0.9\textwidth]{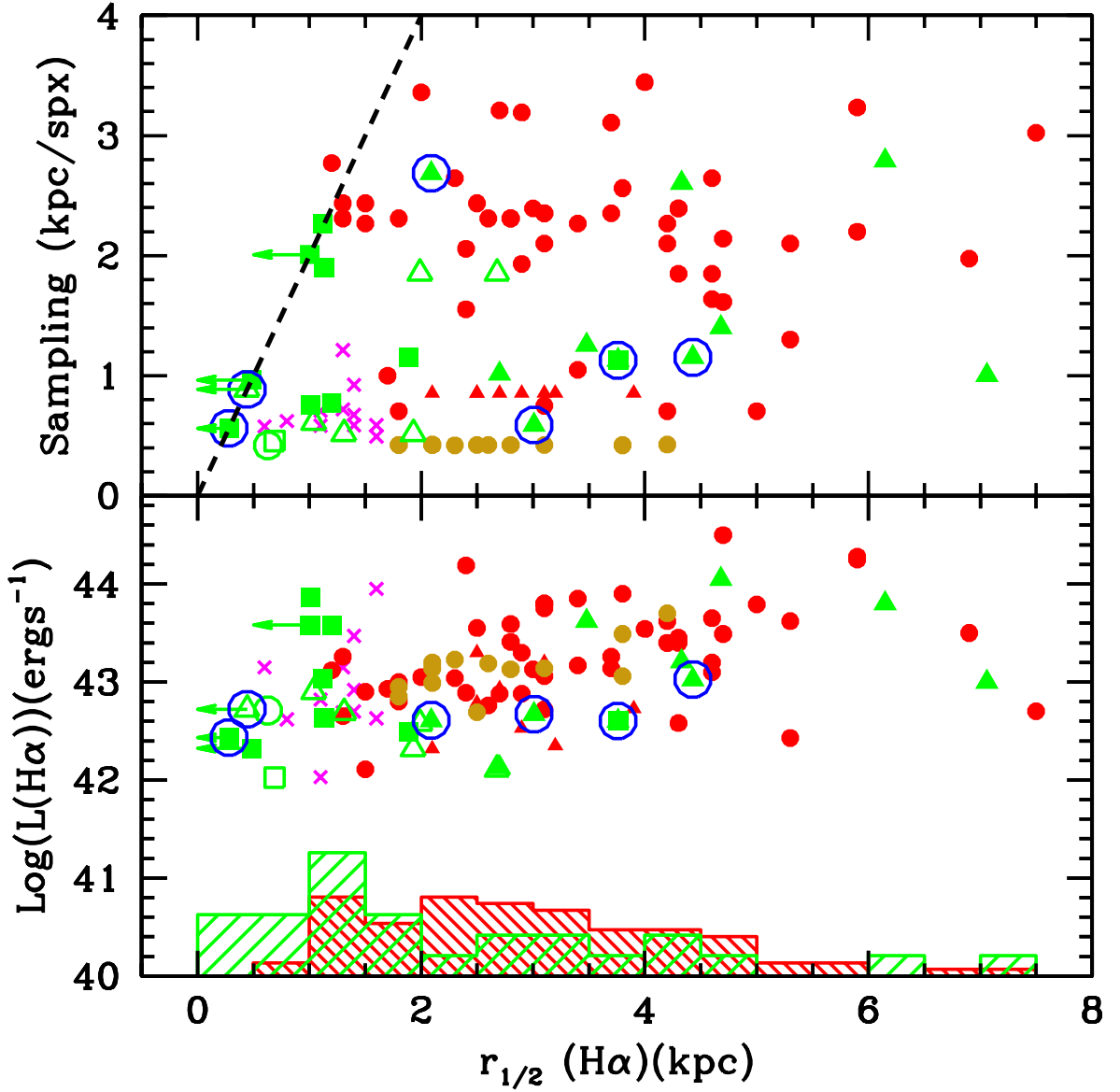} 
 \caption{ Deconvolved half-light radius derived using the Curve of Growth method as a function of the reddening corrected H$\alpha$ luminosity (lower panel) and the spatial sampling (upper panel). For the local U/LIRG sample (green symbols), we used the same code as in Fig.1.  Only the local U/LIRGs with (reddening corrected) H$\alpha$ luminosities above 10$^{42}$ erg s$^{-1}$ and  observed with a spatial resolution similar to that achieved for  distant galaxies (i.e. sampling $>$ 400 pc spx$^{-1}$) are considered. The high-z SFG samples are represented with the following symbol code: Red circles, SINS sample (Forster-Schreiber et al. 2009); magenta crosses, Law et al (2009);  red triangles, Wright et al. (2009);  orange circles, Wisnioski et al. (2011). Typical errors in r$_{1/2}$ range from 30 to 50 percent (for detailed individual errors see Table 2). The dashed line in the upper panel approximately defines the region out of reach owing to  the limited resolution (i.e. left of the line).  The cases for which the derived radii were equal to or smaller than those of a point source are shown as upper limits. The histograms for the local (green) and high-z (red) samples are normalized to the total number of objects in each sample. \label{Fig8}}
 \end{figure*} 

To select from our sample objects well-suited to a proper comparison, we imposed the following two conditions: that i) their linear resolution, and ii) their  (dereddened) H$\alpha$ luminosity should be within the range of the high-z samples. Regarding the spatial resolution, we restricted the local sample to galaxies observed on spatial scales larger than 400 pc / pixel. This physical scale is equivalent to that obtained with a 50 mas /pixel at z $\sim$ 2 (i.e. the finest plate scale used so far). The second criterion restricts the local comparison sample to galaxies with L(H$\alpha$) $>$ 10$^{42}$ ergs$^{-1}$.  After imposing these two conditions, we ended up with a comparison sample of 26 local systems. Eighteen objects are ULIRGs, and six high-luminosity LIRGs (i.e. log(L$_{ir}$/L$_\odot$)$>$11.5). \footnote {The remaining two are the galaxies C and S in the interacting system  IRAS F06259-4708, for which no accurate individual  L$_{ir}$  values exist.  However,  taking into account the infrared luminosity of the whole system log(L$_{ir}$/L$_\odot$ = 11.91), and their H$\alpha$ luminosities, they are likely LIRGs.} Therefore, the comparison sample is formed predominantly by our most luminous systems, with a majority of ULIRGs. Six objects have evidence of hosting a weak AGN based on of their  optical spectra (see Table 1).\footnote{IRAS F07027-6011N, which meets the luminosity and resolution criteria, was not included in this sample as it has an H$\alpha$ luminosity that is anomalously high (2.92 $\times$10$^{42}$ ergs$^{-1}$) for its relatively low infrared luminosity (log (L$_{ir}$/L$_\odot$)= 11.02). This, together with a broad H$\alpha$ emission line and a high [NII]/H$\alpha$ ratio, is a clear indication of significant AGN contamination.} 
 
As for the high-z samples, we selected only sources observed with IFS in H$\alpha$ under good angular resolution conditions (i.e.  with an average FWHM (PSF) of $\le$ 0.6 arcsec -- 5 kpc at z $\sim$ 2). These include the SINS sample  (F{\"o}rster Schreiber et al. 2009, -- hereafter FS09--, and references therein),  and those observed  by Law et al. (2009), Wright et al. (2009), and  Wisnioski et al. (2011).  
The SINS sample consists of 62 objects at z $\sim$ 2 detected in H$\alpha$ with SINFONI (Eisenhauer et al. 2003) at the VLT.  Most of the sample consists of infrared-selected objects covering the mass range 2$\times$10$^{9}$ M$_\odot$ $<$ M$_\star$ $<$ 3.2$\times$10$^{11}$ M$_\odot$, and it  is considered to be a good representation of massive (M$_\star$ $\gtrsim$ 10$^{10}$ M$_\odot$) actively star-forming galaxies at that redshift (FS09).  Several sources were observed with the AO system. 
Wright et al. (2009) observed 6 SFGs at 1.5 $<$ z $<$ 1.7 at the Keck II telescope using the near-infrared integral field spectrograph OSIRIS (Larkin et al. 2006) on the LGS-AO 0.1$''$ lenslet scale.  The objects were selected from the rest-frame UV color-selected catalog of Steidel et al. (2004), and cover the stellar mass range 2$\times$10$^{9}$ - 1.6$\times$10$^{10}$ M$_\odot$. Using the same instrument Law et al. (2009) detected H$\alpha$ emission in 11 SFGs at 2.00 $<$ z $<$ 2.42 from a larger observing sample. With stellar masses in the range 10$^9$ - 8$\times$10$^{10}$ M$_\odot$, these galaxies are considered to be representative of the mean stellar mass of star-forming galaxies at similar redshifts. The selected spaxel scale was 50 mas but to increase the S/N, the data-cubes were smoothed with a Gaussian of a typical FWHM of 150 mas. Using the same instrument,  Wisnioski et al. 2011 obtained AO-assisted IFS data for a sample of 13 SFGs from the UV-selected WiggleZ Dark Energy Survey, with strong [OII] emission lines and 1.28 $<$ z $<$ 1.46. Their stellar masses are in the range 6.3$\times$10$^{9}$ - 5$\times$10$^{11}$M$_\odot$, so are at the high end of the stellar mass distribution probed by IFS samples.  \footnote{Despite the interest of SMGs galaxies to the study U/LIRGs (e.g. Tacconi et al. 2006, 2008; Swinbank et al. 2004), there are very few IFS H$\alpha$ data sets for these objects (though see Tecza et al. 2004; Nesvadba et al.  2007; Swinbank et al. 2006). Unfortunately, the AGN contamination and the lack of H$\alpha$ half-light radius determinations in most cases, prevent us from making a direct and homogeneous size comparison for this class.}  As a whole, these samples should therefore be representative of the intermediate - to-massive SFGs at z $\sim$ 2. 

In Figure 3, we compare the intrinsic (deconvolved) H$\alpha$ sizes obtained for the local U/LIRGs with those for the high-z samples from the literature. We note that this is a rather homogeneous and direct comparison as it is done using the same tracer (i.e. H$\alpha$),  the same technique (i.e. IFS), and similar linear resolutions (i.e. $\sim$ one to few kpc scale). In addition, we used the same method to infer half-light radii as that used for most of high-z determinations (i.e. Curve-of-Growth).  In the bottom panel, the extinction-corrected L(H$\alpha$) \footnote {For the high-z samples, Av values are obtained from the literature SED fittings to broad-band global magnitudes (typically, UV, optical and near-IR, and, in some cases, IRAC and MIPS data. For further details, see references below). These Av(SED) were  transformed to Av(nebular) following Calzetti et al. (2000), who measured E$_{stellar}$(B-V) = (0.44 $\pm$ 0,03)E$_{nebular}$(B-V)  (i.e. Av(nebular)=Av(SED)/0.44). Therefore, for those works that used Av(SED) to correct the observed luminosities  (Wright et al. 2009, Law et al. 2009, Wisnioski et al 2011), this correction was recalculated by considering Av(nebular). For the local sample, Av values are mainly derived from the Balmer decrement within apertures of about 1-3 kpc (see GM09, RZ11).  For most of the VIMOS sample, the H$\alpha$ luminosities were obtained directly from the fluxes given by RZ11. For the objects that were not corrected for reddening in RZ11 owing to the lack of measurements of their Balmer decrement, we took E(B-V)=0.9. This is the mean value obtained for the rest of the sample, and agrees well with the one obtained by Veilleux et al. (1999) for their sample of ULIRGs  (using an extraction aperture of 2 kpc). For the INTEGRAL sample, we took the extinction values from GM09, which were calculated within typical apertures of 2-3kpc. That both, Av(nebular) and Av(SED) values are based on  flux-weighted measurements reduces the effects of the different physical scales used.} is shown as a function of  r$_{1/2}$ (H$\alpha$). To discuss potential biases caused by the limited angular resolution, we plot in the top panel of the figure the radii as a function of the spatial scale of the resolution element used in the different observations. This corresponds to the linear coverage on target of one spaxel or half the FWHM of the PSF, whichever dominates the actual resolution. \footnote {For our local IFS data we took the spaxel size, for Wright et al 2009 data we consider 0.1 arcsec, for Law et al. 2009 data we consider half the FWHM of the PSF after smoothing (see their Table 1), for Wisnioski et al. 2011 data 0.05 arcsec, and for SINS data half of the PSF FWHM, and for objects without seeing measurements a mean value of 0.275 arcsec. }  We note that this is merely an indication of the actual spatial resolution, which also depends on the changing seeing conditions. In the top panel, we also plot the region out of reach owing to the limited angular resolution (i.e. left of the dashed line). \footnote {Due to its approximative nature some values can appear in the {\it forbidden} region.}  

From the comparison of distant massive SFGs and local U/LIRG in Figure 3, we can draw several results (also see Table 1).
 
First, local U/LIRGs have a similar range of sizes as high-z galaxies. Some local U/LIRGs behave in this plot as the small compact objects observed by Law et al (2009), while the rest covers the area defined by the other high-z samples.  

Second, according to the size histograms (bottom panel), small objects (i.e. r$_{1/2}$ $<$ 1 kpc) exist with a higher frequency in the local sample of U/LIRGs.\footnote{ A Kolmogorov-Smirnoff test indicates that the two size distributions do not match with a probability of  99 $\%$ or more.}  In principle, one could think that this is a selection effect since such compact objects cannot be resolved with SINFONI under seeing-limited conditions, and this is the observing set-up for the majority of the high-z galaxies in the plot.  As indicated by the dashed line in the top panel of the figure, small distant galaxies cannot be clearly probed in  typical seeing resolutions. Interestingly, while we find four unresolved objects in the local sample, all the high-z galaxies can be resolved. \footnote{ Note that SINS objects D3a--7429 and GMASS--1146, for which only an upper limit to their sizes were reported by FS09, were observed under relatively bad seeing conditions, hence implying a spatial sampling not probed by the plot.} All this supports the result that galaxies with compact H$\alpha$ emitting regions, such as those found locally, are less frequent at high-z.  The distributions indeed show that  while at high-z fewer than 3 percent (2 out of 81) have r$_{1/2}$$<$ 1 kpc, and fewer than 25 percent (20 / 81) have r$_{1/2}$$<$ 2 kpc,  much higher percentages of 23  $\%$ (6/26) and 58 $\%$ (15/26), respectively, are found for the local sample.  We note that the difference between the local and distant galaxy size distributions would be even clearer if the results of Law et al. 2009 were biased towards small radii as a consequence of the low sensitivity of the instrumental configuration (see Table 1). This possibility has been suggested (Law et al. 2009, FS09), but it is still unclear whether the sizes found could be the result of the intrinsic properties of these galaxies, which, on average, are somewhat less massive than the other samples.  

We now check for possible systematic effects introduced by AGNs (see Sec. 3.3). In Fig. 2,  we observe that the systems with evidence of hosting an (optically weak) AGN \footnote {Note that those objects with evidence that a strong AGN affects the size determination were removed from the comparison sample.} do not affect the general size distribution (also see Table 1). High-z samples also contain known AGNs (e.g. SINS, FS09; Shapiro et al. 2009). However, the five AGNs in the SINS sample (Q1623-BX663, K20-ID5, D3a-7144, D3a-15504, ZC-1101592) have normal sizes with r$_{1/2}$(H$\alpha$) ranging from 3.4 kpc to 5.0 kpc (see their Table 6), hence the effects of the AGNs in this comparison are estimated to be small.

Another interesting result drawn from the figure is that for a sizable fraction of local U/LIRGs, H$\alpha$ emitting regions with large sizes, similar to those of many high-z objects, do exist. In fact, 30 $\%$ (8 out of 26) of the objects in the present local sample have half-light radii larger than 3 kpc. All of these systems are ULIRGs in a pre-coalescence merger phase,  as indicated in Section 3.1.  None is a single nucleus object, either a disk  or a post-merger. However, we note that these objects were classified as pre-coalescence mergers thanks to relatively deep broad-band continuum imaging obtained from the ground and, in most of the cases, with HST as well.  In general, there is no imaging of similar quality for the high-z samples, hence their structure and dynamical phase are not known with a similar level of detail.  However, F{\"o}rster-Schreiber et al. (2011) obtained  deep rest-frame (continuum) optical high-angular resolution HST NICMOS imaging for 6 SINS objects, and compared it with the H$\alpha$ emission from their SINFONI IFS data. One of these objects (BX528) had been previously classified as a merger based on its kinematics (FS09), and has a double-nucleus irregular continuum emission (see their Fig. 1),  with an associated total radius of  4.86 kpc (or  3.18 kpc and 3.57 kpc for the individual components). This object has an r$_{1/2}$ (H$\alpha$) of 4.6 kpc, so it has the typical characteristics of one of the pre-coalescence local U/LIRGs.  The structure of the H$\alpha$ emission for high-z samples (e.g. FS09) often shows  a clumpy and irregular structure similar to that found in the  H$\alpha$ maps of local pre-coalescence U/LIRGs. Furthermore, on the basis of kinematic arguments FS09 found that 33 $\%$ of the SINS galaxies have evidence of being interacting and merging systems. Kartaltepe et al. (2011) also found a significant fraction of mergers, interacting, and irregular galaxies at z$\sim$ 2, especially among ULIRGs. Therefore, it is likely that pre-coalescence systems similar to those in the local sample also exist in the high-z samples. Interestingly, the median properties of the whole high-z sample are similar to those of pre-coalescence local U/LIRG systems (Table 1). 
If the pre-coalescence systems are excluded from the local U/LIRGs sample, the high-z SF galaxies appear to be larger than the local U/LIRGs by a factor of about 3 (see Table 1 for specific values).

The depth of the observations is another factor that can affect the comparison between the local and distant samples.  It is difficult to estimate the depth of the different observations owing to the many factors affecting its calculation, some of which are difficult to handle (e.g. seeing fluctuations, wavelength-dependent infrared background). However, if the observed integrated H$\alpha$ fluxes of the local U/LIRGs are transformed into surface brightness and redshifted to z=2.2, there are several cases with values  below the  $\sim$ 1.$\times$10$^{-17}$ erg s$^{-1}$ found by FS09 for the faintest sources of the SINS sample. The objects below this threshold are generally large objects of low surface brightness, which would have been missed  if they were at z $\sim$ 2.  Therefore,  the relatively shallower observations of the high-z samples strengthen the conclusion that local U/LIRGs are on average intrinsically smaller than high-z populations.

\subsection{H$\alpha$ and SF surface densities in star-forming galaxies at low and high redshifts}

 \begin{figure}[h]
 \centering  

 \includegraphics[width=8cm]{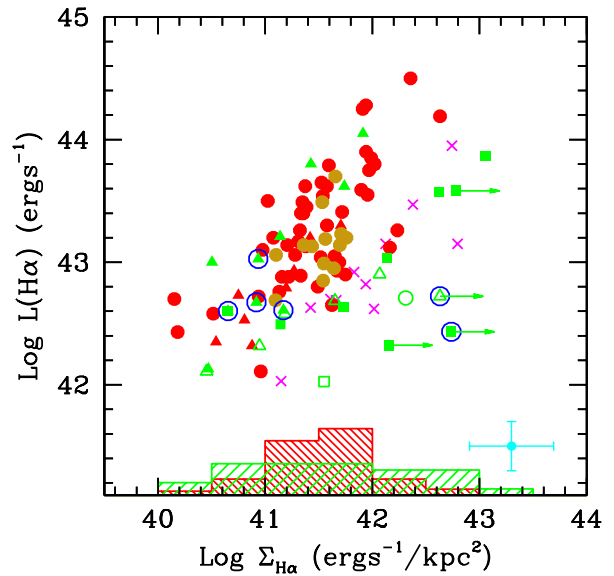}
 \caption{ H$\alpha$ luminosity density as a function of the H$\alpha$ luminosity for local U/LIRGs and distant massive SFG with a similar luminosity range and observed under similar linear resolution conditions. Symbols are the same as in Fig. 3. The histograms for the local (green) and high-z (red) samples are normalized to the total number of objects in each sample. The blue cross in the lower-right corner indicates the typical errors.}
\
 \label{Fig2}
 \end{figure}

Recent works (e.g. Iono et al. 2009, Rujopakarn et al. 2011) have suggested that the overall size of the star-forming region in local U/LIRGs is significantly smaller than in both high-z U/LIRGs and SMGs, and local lower-luminosity star-forming galaxies.  Rujopakarn et al. (2011) suggested that the star formation in local U/LIRGs proceeds with extremely high SF surface densities, in some cases by factors of 1000 or higher than in other SFGs, and therefore that they are driven by a unique process. These results are based (1) on the determination of the physical sizes using different spectral features for  the local and high-z star-forming galaxies, and (2) on the assumption that these different features give a similar measure of the size of the SF region.  The available spectral features  trace different physical mechanisms and components of the ISM. The most commonly used tracers are the  H$\alpha$ and Pa$\alpha$ lines (warm ionized gas), CO lines (cold molecular gas), 24$\mu$m continuum emission (hot dust emission), and radio continuum (non-thermal synchrotron emission). Each of these tracers is affected by its own biases (see discussion in Genzel et al. 2010) hence an effort should be made to establish the size measurements, and therefore the SF surface density, in a homogeneous way when comparing galaxy samples.

To compare our local U/LIRGs with high-z samples, we converted the extinction-corrected L(H$\alpha$) luminosities into H$\alpha$ luminosity surface densities within r$_{1/2}$  ($\Sigma_{H\alpha}$= L(H$\alpha$)/2$\pi$r$_{1/2}^2$) using the H$\alpha$ half-light radius derived from the same IFS data (Section 3.1).  Therefore, both the half-light radius and the luminosity were inferred from the same set of data, namely the IFS-based H$\alpha$ emission maps.  In Figure 4, we make a direct and homogeneous comparison between our local U/LIRGs and  high-z massive SF galaxies covering the same H$\alpha$ luminosity range, and with similar IFS H$\alpha$ measurements available from the literature (see Section 4.1; Table 1).

Local U/LIRGs and high-z SF galaxies cover a similar range in H$\alpha$ luminosity surface density spanning over at least three orders of magnitudes from $\Sigma_{H\alpha}$  of 10$^{40}$ to more than 10$^{43}$ erg s$^{-1}$ kpc$^{-2}$. This would correspond  to  $\Sigma_{SFR}$ between 0.08 and 80 M$_{\sun}$ yr$^{-1}$ kpc$^{-2}$ according to Kennicutt (1998). However, while local U/LIRGs are more evenly distributed towards high surface density values (i.e. $\Sigma_{H\alpha}>$ 10$^{42}$ erg s$^{-1}$ kpc$^{-2}$), the majority of high-z galaxies have moderate surface densities (i.e. $\Sigma_{H\alpha}$ between 10$^{41}$ and 10$^{42}$ erg s$^{-1}$ kpc$^{-2}$). 

The data in Fig. 4 are also consistent with a trend in the sense that the more luminous  galaxies also have higher luminosity surface densities. This linear behavior is more clearly observed in the the high-z sample (correlation coefficient $=$ 0.60)  than locally (0.43). The local galaxies with  H$\alpha$ surface densities that are significantly higher than the high-z sample (i.e.  $>$ 2$\sigma$ from the mean linear fit) represent about  20$\%$ of the sample. If the Law et al. (2009) sample (indicated by crosses in Fig. 4) is not considered within the high-z pool a stronger correlation is obtained (0.72), and the percentage of local galaxies that depart  significantly from the mean high-z behavior increases by up to about 40$\%$. 

Only two (optically) identified AGNs are among the local U/LIRGs with high H$\alpha$ surface densities that are therefore suspected to be contaminated. However, the majority of AGNs are in sources with low surface densities, suggesting that they cannot be the dominant factor for the local high surface densities. 
Therefore, although the presence of a buried AGN cannot be discarded (see Section 3.3), the H$\alpha$ emission is likely due mainly to star formation. 

The trend shown in Fig. 4 is equivalent to the one presented by Rujopakarn et al. (2011), based on the total infrared (TIR) luminosity (i.e. $\Sigma_{TIR}$ vs. L$_{TIR}$, see their figure 4).   However,  while in their TIR-based relation most of the the local U/LIRGs largely depart from the general trend found in other (local and distant) populations, we do not observe this behavior in H$\alpha$, except in some cases.  More specifically, while Rujopakarn et al. (2011) found that most of the local U/LIRGs depart by factors of between 1 and 3 orders of magnitude in luminosity surface density away from other the trend defined by other SFGs samples, our H$\alpha$ based measurements suggests that only a fraction (20-40 $\%$) show a clear departure of about one order of magnitude, as can be inferred from Fig. 4.

\subsection{Dependence of size on Av and [NII]/H$\alpha$: Comparison of local U/LIRGs with infrared-selected SINS galaxies}

As discussed above, the reddening structure may affect the derived sizes for the H$\alpha$ emitting region. Therefore, the effects of a possible different radial variation in the extinction for local and high-z sources should be in principle taken into account when comparing both samples.  Unfortunately our knowledge of the 2D dust / reddening structure in ULIRGs is mostly unknown (see GM09b, and Sec. 3.1), and we have a complete ignorance of it for high-z samples. Therefore a detailed study of this topic is well beyond the current possibilities. 

As a first attempt at studying relative effects, we compare the Av distributions of the samples under analysis, looking for possible correlations with the derived sizes (Fig. 5).  No obvious correlation size - Av is found for any of the two samples. We note that the lack of this correlation does not imply that reddening does not  affect size, as it is the radial variation of extinction, rather than its global value, the main contributing factor.
  
Fig. 5 also shows that there is a clear distinction in the behavior of local and distant Av distributions. In particular, while the local sample show a large scatter with values in the range 0.5 - 6 mag, the high-z sample distribution has a lower mean value with a peak at about 1.2 mag, hence on average the high-z sample is less extincted. However, we see below that a sub-sample of high-z targets (i.e.  those that are IR-selected) has a distribution of Av similar to the local sample.

 \begin{figure}[h]
 \centering  

 \includegraphics[width=8cm]{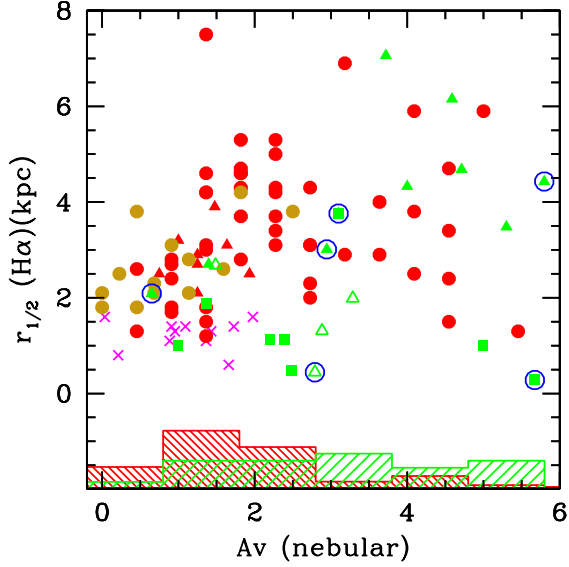}
 \caption{ Av (nebular) as a function of H$\alpha$ radius.  Symbol code is the same as in previous figures.  The histograms for the local and high-z samples are normalized to the total number of objects in each sample. For the high-z samples, Av were  obtained from the literature SED fittings to global magnitudes,  transformed to Av(nebular) following Calzetti et al. (2000) (i.e. Av(nebular)=Av(SED)/0.44). For the local sample, Av values were derived from the Balmer decrement within apertures of about 1-3 kpc (see GM09, RZ11).} 

\
 \label{Fig4}
 \end{figure}

 \begin{figure}[h]
 \centering  

 \includegraphics[width=8cm]{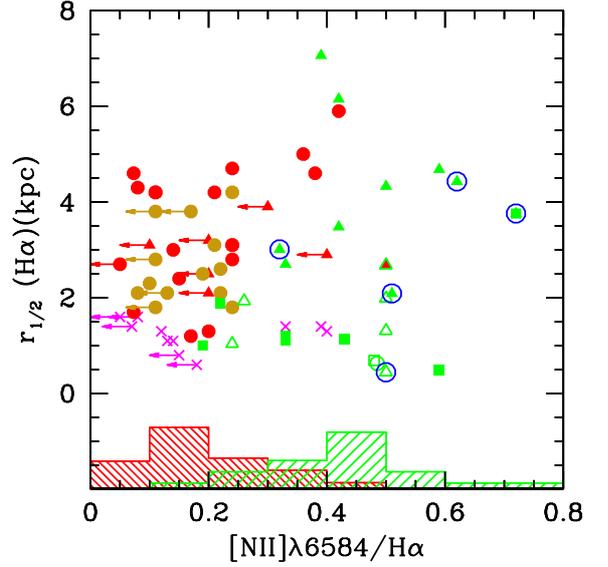}
 \caption{ [NII]/H$\alpha$ as a function of H$\alpha$ radius (see text). Symbol code is the same as in previous figures.  The histograms for the local and high-z samples are normalized to the total number of objects in each sample.}

\
 \label{Fig5}
 \end{figure}

 \begin{figure}[h]
 \centering  

 \includegraphics[width=8cm]{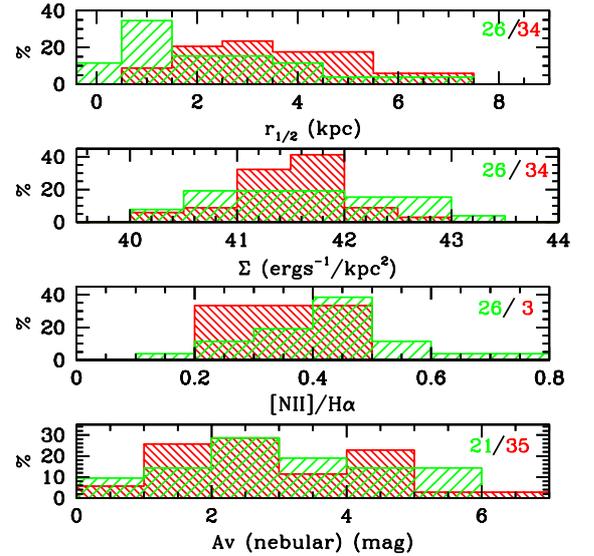}
 \caption{Comparison of different properties between the local U/LIRGs and the infrared-selected SINS objects from F{\"o}rster-Schreiber et al (2009).   The histograms are normalized to the total number of values in each sample, which are indicated in the upper- right corner of each panel. Note that the metallicity distribution for the high-z sample is defined by only three values. }

\
 \label{Fig4}   
 \end{figure}

For the [NII]/H$\alpha$ ratio, there is also a clear difference between the local and distant samples, as shown in Fig. 6. \footnote{For the INTEGRAL data, we use the integrated values for the circumnuclear region given by Garc{\'{\i}}a-Mar{\'{\i}}n (2007), while for the VIMOS data we consider the integrated spectra following Rosales-Ortega et al. (2011). Hence, typical physical scales are a few kpc, which are similar to the high-z samples. We note that these are flux-weighted measurements.} Similarly to Av, there is again no correlation between this ratio and size. The difference between the two  [NII]/H$\alpha$  distributions may be even larger than suggested by the histograms, as several values for the high-z sample are only upper limits, as indicated by the arrows in the figure. \footnote {If we interpret the [NII]/H$\alpha$ ratio as a measure of metallicity via the N2 index (e.g. Pettini and Pagel, 2004), we obtain a mean solar value of 12 + log (O/H) = 8.70 (Allende-Prieto et al. 2001; Asplund et al. 2004; Scott et al. 2009) for the present local sample. This agrees within the uncertainties with the findings of Rupke et al. (2008), who derive metallicities from the nuclear spectra of a local sample of U/LIRGs using different calibrators. For the high-z sample  as a whole, an upper limit of 8.43 is obtained, so at least a factor of two lower metallicity with respect to the solar, in agreement with the previous long-slit results of Erb et al. (2006b).  We should note that, this ratio might also be affected by the presence of an AGN and by shock excitation as found in local (e.g. Veilleux et al. 1995, 1999;  Yuan et al. 2010;  Monreal Ibero et al. 2006, 2010)  and distant (Shapiro et al. 2009) samples. However, according to the diagnostic diagram proposed by Veilleux and Osterbrock (1987), the [NII]/H$\alpha$ values reported in figure 6 suggest that ionization is mainly due to stars (i.e. [NII]/H$\alpha$ $<$ 0.6), for a wide range of [OIII]/H$\alpha$ values.} However, it should be noted that [NII]/H$\alpha$ values are available for only a fraction of the SINS sample  and these are strongly biased towards the BX (i.e. optically-selected ) objects (F{\"o}rster Schreiber et al. 2006). 

The differences between the mean Av and [NII]/H$\alpha$ values of the local and distant samples shown above are dominated by the contribution of the UV-selected high-z galaxies clustered around the low Av values. If we restrict the comparison to the infrared-selected galaxies of the SINS sample (see table 1 of FS09), the A$_v$ (nebular) distribution is similar to that of the local sample ( $<$A$_v$$>$=3.2 $\pm$ 1.5 in both cases), as it can be seen in the lower panel of figure 7.  For the three infrared-selected / bright SINS objects with [NII]/H$\alpha$ measurements (D3a-6004, Da3-15504, ZC-782941; FS09, Genzel et al. 2008), a mean value of [NII]/H$\alpha$ = 0.34 was found, which is close to the value found for the local U/LIRGs and significantly higher than those for the rest of the high-z galaxies.  In Figure 7, we also compare the size and luminosity surface density of the local U/LIRGs with those of the infrared-selected objects in the SINS sample.  Despite the similarity in the mean Av and [NII]/H$\alpha$ properties, the size distributions are different, with a higher frequency of local objects of small size (high surface density). Although the number of objects is low for a reliable statistics, a Kolmogorov-Smirnoff test indicates that the radius distributions do not match with a probability of 99$\%$ or more.

\section{ Summary and conclusions} 

Using IFS-based H$\alpha$ emission maps, we have derived the sizes of the SF regions for a representative sample of 54 local U/LIRGs (66 galaxies).  From this initial sample we have selected a sub-sample of 26 local U/LIRGs which we have compared with existing works of high-z SFGs observed in H$\alpha$ with near-infrared IFS at high angular resolution (i.e. AO-assisted or under good seeing conditions) to match the physical size probed locally and at high-z.  We have also selected these galaxies to have similar  L(H$\alpha$) (i.e. SFRs) as high-z samples. The comparison between the basic properties (such as size and luminosity surface density) of the galaxies in the local and distant samples has been made therefore in a homogeneous way, using the same observing technique (IFS), tracer (H$\alpha$), and similar linear resolutions.  The main conclusions of the present analysis can be summarized as follows:

1- The SF region sizes for the whole local sample of U/LIRGs, as derived from H$\alpha$,  have a wide range of values with r$_{1/2}$(H$\alpha$)  from 0.2  to  7 kpc. For objects with L$_{ir} >$ 10$^{11.4}$ L$_\odot$, about 2/3 have relatively compact (i.e.  r$_{1/2}$ $<$ 2 kpc) emission. For the remaining 1/3 (13/41) with large H$\alpha$ emitting regions (i.e.  r$_{1/2}$ $>$ 2 kpc), 11 are ULIRG in a pre-coalescence merger phase with nuclear separations between 1.5 and 14 kpc. For objects with L$_{ir} < $ 10$^{11.4}$ L$_\odot$, large H$\alpha$ emission is not necessarily associated to mergers, but to isolated disks as well. 

When we compare the subsample of U/LIRGs with the high-z sample (at similar L(H$\alpha$) and physical scales) we find the following additional conclusions:   
   
2- The H$\alpha$ size distributions for  local U/LIRGs and distant massive star forming-galaxies cover a similar range of values. However, we have found that there is a higher frequency of compact objects locally than at high-z. In particular, while objects with r$_{1/2}$ $<$ 2 kpc represent  58  $\%$  (15/26) of the low-z subsample, they constitute fewer than 25 $\%$  (20/81) of the galaxies at high-z.  The median size of local U/LIRGs is a factor of 1.5 smaller than the one for high-z SFGs. This value strongly depends on whether only pre-coalescence systems are considered ($\sim$1) or  are excluded ($\sim$ 3) from the local U/LIRG subsample. These factors are on average smaller than the ones reported using a variety of other tracers.     

3- Most of the local U/LIRGs and high-z massive SF galaxies cover a similar region in the $\Sigma_{H\alpha}$ - L( H$\alpha$) plane,  spanning from $\Sigma_{H\alpha}$  of 10$^{40}$ to more than 10$^{43}$ erg s$^{-1}$ kpc$^{-2}$.   About 20-40 $\%$ of the local U/LIRGs studied here show values of  $\Sigma_{H\alpha}$ significantly higher (by factors of  about 10)  than at high-z.   These are considerably smaller than the much higher factors of  $\sim$ 1000 or more, recently reported in similar planes (i.e. L(TIR) vs.  $\Sigma_{TIR}$).

4- When the comparison with the high-z SFGs is restricted to the infrared-selected SINS objects,  local and distant samples  have similar properties in terms of global visual extinction (and  [NII]/H$\alpha$). Although with a lower statistical significance,  the main result of  the local vs. high-z comparison remains unchanged:  the SF region size, on average,  is a distinctive factor between local U/LIRGs and high-z samples of similar L(H$\alpha$) (i.e. SFR).  

5- A significant fraction (approximately 1/3) of local U/LIRGs, generally pre-coalescence merger systems, are indistinguishable from typical high-z SFGs galaxies in terms of their H$\alpha$ size and surface brightness.   

\onecolumn

\begin{longtable}{lccccc}
\caption{H$\alpha$ radii for the IFS sample of U/LIRGs observed with VIMOS and INTEGRAL} 
\label{table:tabla2}
\\
\hline

IRAS ID & Other ID & Subsample &  r$_{1/2}$-H$\alpha$ (A/2)     & r$_{1/2}$-H$\alpha$ (CoG)  &   Comment \\
name    &    	   &    &     (kpc)              &     (kpc)          &           \\
 (1)    &  (2)     &(3) &      (4)               &      (5)           &     (6) \\

\hline

\endfirsthead
\multicolumn{6}{c}{{Table~\ref{table:tabla2} -- Continued from previous page}} \\

\hline

IRAS ID & Other ID & Subsample  &  r$_{1/2}$-H$\alpha$ (A/2)     & r$_{1/2}$-H$\alpha$ (CoG)  &   Comment \\
name    &    	   &    &     (kpc)              &     (kpc)          &           \\
 (1)    &  (2)     &(3) &      (4)               &      (5)           &     (6) \\

\hline
\endhead	
\hline
\multicolumn{6}{c}{{Continues on next page}}\\
\endfoot
\hline \hline
\noalign{\smallskip}

\multicolumn{6}{@{} p{0.99\columnwidth} @{}}{\scriptsize \textbf{Notes.} Columns: (1) and (2) Identification, (3) Indicates whether or not the object is in the subsample of U/LIRGs with L(H$\alpha$) and spatial resolution similar to those of high-z samples (see text)  (4) H$\alpha$ radius obtained with the A/2 method, (5)  H$\alpha$ radius obtained with the Curve-of-Growth method, (6) comments with the following code: 0:H$\alpha$ half-light radius equal or smaller than the PSF (i.e. unresolved), 1: H$\alpha$ half-light radius smaller than 1.25 times the radius of the PSF and therefore, with larger uncertainty, 2: Noisy continuum,  3: Limited FoV, 4: Observed with INTEGRAL / SB1 configuration, 5: Evidence for an AGN according to the emission line ratios (see GM09, and compilation of RZ11), 6: Evidence for a broad line associated to H$\alpha$ suggesting the presence of an AGN. 7: The L$_{ir}$ for this object is unknown, and therefore it is not included in the analysis. 8: For this object the individual L$_{ir}$ is not known, but taking into account that the whole system has L$_{ir}$=11.91, and (reddening corrected) H$\alpha$ luminosities above 10$^{42}$ erg s$^{-1}$ it is likely a LIRG, and it kept for the analysis. 9: For this object the individual L$_{ir}$ is not accurate known, but taking into account that the whole system has L$_{ir}$=11.91 and according to Spitzer MIPS (24$\mu$m) data it carries a large fraction of the IR flux, it is likely a LIRG. 10: This object likely to have a strong AGN contamination so it was not included in the sample for comparison with high-z samples (see text). The values reported for the CoG and A/2 methods have been {\it deconvolved} in quadrature from the radii associated to stars. We used an averaged value of r$_{1/2}$ (PSF) = 1.07 arcsec for VIMOS data from 7 stars located within the FoV of the science  pointings. For the INTEGRAL-SB2 data,  that value was scaled according to the spaxel scale, and for SB1 (which was used on average under better seeing conditions) we adopted 0.9 arcsec, as inferred from the Mrk231 observations. Typical errors are estimated to be 30 percent, except for cases with comments 1and  2 for which are increased to 50 percent. Values  limited by  FoV  (comment 3) should be considered as lower limits.  For radii derived from the INTEGRAL data, typical errors are estimated to be somewhat larger (40 percent), since no direct measurements for the PSF were available. Further details on these galaxies can be found in GM06 (INTEGRAL data)  and RZ11 (VIMOS data).\newline $^a$ Identification as in Garc{\'{\i}}a-Mar{\'{\i}}n et al.(2006).}

\endlastfoot
\hline
& &VIMOS sample& & &\\ 
\hline
F01159$-$4443N &   ESO$-$244$-$G012       &  N  &  0.35$\pm$   0.10 &	0.36$\pm$    0.11 &	 1,2\\
F01341$-$3735N &   ESO$-$297$-$G011       &  N  & $>$1.36$\pm$ 0.41 &$>$1.89$\pm$    0.57 &	 2,3\\
F01341$-$3735S &   ESO$-$297$-$G012       &  N  &  0.23$\pm$   0.11 &	0.22$\pm$    0.11 &	 1  \\
F04315$-$0840  &   NGC~1614		  &  N  &  0.51$\pm$   0.15 &	0.51$\pm$    0.15 &	    \\
F05189$-$2524  &			  &  Y  &    $<$0.28	    &	  $<$0.28	  &	 0,5,6\\
F06035$-$7102  &			  &  Y  &  3.72$\pm$   1.12 &	7.06$\pm$    2.12 &	    \\
F06076$-$2139N &			  &  N  &    $<$0.25	    &	  $<$0.25	  &	 0  \\
F06076$-$2139S &		          &  N  &  0.48$\pm$   0.24 &	0.45$\pm$    0.22 &	 1,7  \\
F06206$-$6315  &		          &  Y &  2.53$\pm$   0.76 &	4.43$\pm$    1.33 &	 5  \\
F06259$-$4708N &		          &  N  &  0.71$\pm$   0.21 &	0.74$\pm$    0.22 &	 9\\
F06259$-$4708C &		          &  Y  &  1.16$\pm$   0.35 &	1.31$\pm$    0.39 &	 8\\
F06259$-$4708S &		          &  Y  &  1.45$\pm$   0.44 &	1.93$\pm$    0.58 &	 8\\
F06295$-$1735  & ESO$-$557$-$G002         &  N  &  1.91$\pm$   0.57 &	2.46$\pm$    0.74 &	 2\\
F06592$-$6313  &		          &  N  &  0.35$\pm$   0.17 &	0.39$\pm$    0.20 &	 1\\
F07027$-$6011N &  AM~0702$-$601           &  N  &  0.47$\pm$   0.24 &	0.50$\pm$    0.25 &	 1,5,6,10\\
F07027$-$6011S &		          &  Y  &  0.63$\pm$   0.19 &	0.63$\pm$    0.19 &	 \\
F07160$-$6215  & NGC~2369	          &  N  & $>$0.54$\pm$ 0.16 &$>$0.86$\pm$    0.26 &	 3\\
08355$-$4944   &		          &  N  &  0.59$\pm$   0.18 &	0.64$\pm$    0.19 &	 2\\
F08520$-$6850  & ESO60$-$IG016            &  Y  &  1.02$\pm$   0.31 &	1.04$\pm$    0.31 &	 2\\
09022$-$3615   &		          &  Y  &  1.30$\pm$   0.39 &	1.20$\pm$    0.36 &	 \\
F09437$+$0317N & IC-563 	          &  N  & $>$2.20$\pm$ 0.66 &$>$3.22$\pm$    0.97 &	 3\\
F09437$+$0317S & IC-564 	          &  N  & $>$ 1.93$\pm$0.58 &$>$3.08$\pm$    0.92 &	 3\\
F10015$-$0614  & NGC-3110	          &  N  & $>$ 1.95$\pm$0.59 &$>$2.66$\pm$    0.80 &	 3\\
F10038$-$3338  & IC2545 	          &  Y  &  0.51$\pm$   0.26 &	0.69$\pm$    0.23 &	 1\\
F10257$-$4339  & NGC~3256	          &  N  & $>$1.01$\pm$ 0.30 &$>$1.37$\pm$    0.41 &	 3\\
F10409$-$4556  & ESO$-$264$-$G036         &  N  & $>$1.78$\pm$ 0.53 &$>$3.24$\pm$    0.97 &	 3\\
F10567$-$4310  & ESO$-$264$-$G057         &  N  & $>$1.61$\pm$ 0.48 &$>$3.69$\pm$    1.11 &	 2,3\\
F11255$-$4120  & ESO$-$319$-$G022         &  N  &  1.25$\pm$   0.38 &	2.70$\pm$    0.81 &	 \\
F11506$-$3851  & ESO$-$320$-$G030         &  N  &  0.68$\pm$   0.21 &	1.05$\pm$    0.31 &	 \\
F12043$-$3140N & ESO$-$440$-$IG 058       &  N  &  0.24$\pm$   0.12 &	0.18$\pm$    0.09 &	 1,7\\
F12043$-$3140S & ESO$-$440$-$IG 058       &  N  & $>$1.38$\pm$ 0.41 &$>$2.92$\pm$    0.88 &	 3\\
F12115$-$4656  & ESO$-$267$-$G030         &  N  &  1.34$\pm$   0.40 &	1.55$\pm$    0.46 &	 \\
12116$-$5615   &		          &  N  &  0.28$\pm$   0.14 &	0.29$\pm$    0.14 &	 1\\
F13001$-$2339  & ESO$-$507$-$G070         &  N  &  0.88$\pm$   0.26 &	0.91$\pm$    0.27 &	 \\
F13229$-$2934  & NGC~5135	          &  N  & $>$0.52$\pm$ 0.16 &$>$0.53$\pm$    0.16 &	 3,5\\
F14544$-$4255E & IC 4518	          &  N  & $>$0.94$\pm$ 0.28 &$>$1.06$\pm$    0.32 &	 3\\
F14544$-$4255W & IC 4518	          &  N  &  0.57$\pm$   0.17 &	0.66$\pm$    0.20 &	 5\\
F17138$-$1017  &		          &  N  &  0.56$\pm$   0.17 &	0.63$\pm$    0.19 &	 \\
F18093$-$5744N &		          &  N  &  1.10$\pm$   0.33 &	1.56$\pm$    0.47 &	 \\
F18093$-$5744C &		          &  N  &  0.35$\pm$   0.11 &	0.34$\pm$    0.10 &	 6\\
F18093$-$5744S &		          &  N  &  0.84$\pm$   0.25 &	1.21$\pm$    0.36 &	 7\\
F21130$-$4446  &		          &  Y  &  1.59$\pm$   0.48 &	1.89$\pm$    0.57 &	 2\\
F21453$-$3511  & NGC~7130	          &  N  & $>$1.04$\pm$ 0.31 &$>$2.27$\pm$    0.68 &	 3,5,6\\
F22132$-$3705  & IC 5179	          &  N  & $>$1.56$\pm$ 0.47 &$>$1.90$\pm$    0.57 &	 3\\
F22491$-$1808  &		          &  N  &  1.70$\pm$   0.50 &	2.30$\pm$    0.70 &	 \\
F23128$-$5919  & AM 2312$-$591            &  Y  &  2.01$\pm$   0.60 &	3.01$\pm$    0.90 &	 5,6\\
\hline
& &INTEGRAL sample& & &\\ 
\hline
06268$+$3509    &  			  &  Y  &  3.74$\pm$   1.50 &	4.33$\pm$    1.73 &	 \\
06487$+$2208    & 			  &  Y  &  0.98$\pm$   0.49 &	1.12$\pm$    0.56 &	 1,4\\
F08572$+$3515   & 			  &  Y  &  1.85$\pm$   0.74 &	2.70$\pm$    1.08 &	 \\
F11087$+$5351   &			  &  Y  &  2.49$\pm$   1.00 &	3.76$\pm$    1.50 &	 4,6\\
		& Arp299E / IC694$^a$	  &  N  & $>$0.68$\pm$ 0.27 &	$>$1.03$\pm$ 0.41 &	 3\\
F11257+5850$^a$ & Arp299W /NGC3690$^a$	  &  N  & $>$0.53$\pm$ 0.21 &	$>$0.67$\pm$ 0.27 &	 3,5\\
F12112$+$0305   &			  &  Y  &  2.60$\pm$   1.04 &	3.48$\pm$    1.39 &	 \\
F12490$-$1009   &			  &  N  &  1.77$\pm$   0.71 &	1.58$\pm$    0.63 &	 \\
F13156$+$0435N  &			  &  Y  &  1.96$\pm$   0.78 &	1.99$\pm$    0.80 &	 \\
F13156$+$0435S  &			  &  Y  &  2.66$\pm$   1.06 &	2.68$\pm$    1.07 &	 \\
F13428$+$5608   & Mrk 273		  &  N  &  1.29$\pm$   0.52 &	1.71$\pm$    0.68 &	 5\\
F13536+1836     & Mrk 463 		  &  Y  &    $<$0.44	    &	  $<$0.44	  &	 0,5,6\\
F14060$+$2919   &			  &  Y  &  1.30$\pm$   0.65 &	1.13$\pm$    0.57 &	 1\\
F14348$-$1447   &			  &  Y  &  3.82$\pm$   1.53 &	4.68$\pm$    1.87 &	 \\
F15206$+$3342   &			  &  Y  &    $<$1.00	    &	  $<$1.00	  &	 0\\
F15250$+$3609   &			  &  Y  &    $<$0.48	    &	  $<$0.48	  &	 0\\
F15327+2340     & Arp 220 		  &  N  &  0.63$\pm$   0.25 &	0.61$\pm$    0.24 &	 \\
F16007$+$3743   & 			  &  Y  &  3.67$\pm$   1.47 &	6.15$\pm$    2.46 &	 \\
F17207$-$0014   & 			  &  Y  &  0.90$\pm$   0.36 &	1.01$\pm$    0.40 &	 \\
F18580$+$6527   &			  &  Y  &  2.37$\pm$   0.95 &	2.09$\pm$    0.84 &	 5\\

\end{longtable}

 \acknowledgements


We acknowledge an anonymous referee for useful comments, which helped us to improve a previous version of the  manuscript.  This paper is based on observations made with the WHT, operated on the island of La Palma by the ING in the Spanish Observatorio del Roque de los Muchachos of the Instituto de Astrof'sica de Canarias.
Also based on observations carried out at the European Southern observatory, Paranal
(Chile), Programs 076.B-0479(A), 078.B-0072(A) and 081.B-0108(A).  This
research made use of the NASA/IPAC Extragalactic Database (NED), which
is operated by the Jet Propulsion Laboratory, California Institute of
Technology, under contract with the National Aeronautics and Space
Administration.  MGM is supported by the German federal department for education and research (BMBF) under the project number 50OS1101. This work has been supported by the Spanish Ministry of Science and
Innovation (MICINN) under grants ESP2007-65475-C02-01 and AYA2010-21161-C02-01. 


\begin{thebibliography}{}
\bibitem[Allende Prieto et al.(2001)]{2001ApJ...556L..63A} Allende Prieto, C., Lambert, D.~L., \& Asplund, M.\ 2001, \apjl, 556, L63
\bibitem[Alonso-Herrero et al.(2009)]{2009A&A...506.1541A} Alonso-Herrero, A., Garc{\'{\i}}a-Mar{\'{\i}}n, M., Monreal-Ibero, A., et al.\ 2009, \aap, 506, 1541 
\bibitem[Alonso-Herrero et al.(2006b)]{2006ApJ...652L..83A} Alonso-Herrero, A., Colina, L., Packham, C., et al.\ 2006b, \apjl, 652, L83 
\bibitem[Alonso-Herrero et al.(2006a)]{2006ApJ...650..835A} Alonso-Herrero,  A., Rieke, G.~H., Rieke, M.~J., et al.\ 2006a, \apj, 650, 835
\bibitem[Alonso-Herrero et al.(2012)]{2012ApJ...744....2A} Alonso-Herrero, A., Pereira-Santaella, M., Rieke, G.~H., \& Rigopoulou, D.\ 2012, \apj, 744, 2 
\bibitem[Arribas et al.(2008)]{2008A&A...479..687A} Arribas, S., Colina, L., Monreal-Ibero, A., Alfonso, J., Garc{\'{\i}}a-Mar{\'{\i}}n, M., \& Alonso-Herrero, A.\ 2008, \aap, 479, 687 
\bibitem[Arribas et al.(2004)]{2004AJ....127.2522A} Arribas, S., Bushouse, H., Lucas, R.~A., Colina, L., \& Borne, K.~D.\ 2004, \aj, 127, 2522 
\bibitem[Arribas et al.(1998)]{1998SPIE.3355..821A} Arribas, S., et al.\ 1998, \procspie, 3355, 821
\bibitem[Asplund et al.(2004)]{2004A&A...417..751A} Asplund, M., Grevesse, N., Sauval, A.~J., Allende Prieto, C., \& Kiselman, D.\ 2004, \aap, 417, 751  
\bibitem[Bouch{\'e} et al.(2007)]{2007ApJ...671..303B} Bouch{\'e}, N., et al.\ 2007, \apj, 671, 303 
\bibitem[Calzetti et al.(2000)]{2000ApJ...533..682C} Calzetti, D., Armus, L., Bohlin, R.~C., Kinney, A.~L., Koornneef, J., \& Storchi-Bergmann, T.\ 2000, \apj, 533, 682
\bibitem[Clements et al.(1996)]{1996MNRAS.279..459C} Clements, D.~L., Sutherland, W.~J., Saunders, W., et al.\ 1996, \mnras, 279, 459 
\bibitem[Colina et al.(2004)]{2004ApJ...602..181C} Colina, L., Arribas, S., \& Clements, D.\ 2004, \apj, 602, 181  
\bibitem[2005]{Col05} Colina, L., Arribas, S., \& Monreal-Ibero, A. 2005, \apj, 621, 725
\bibitem[D{\'{\i}}az-Santos et al.(2011)]{2011ApJ...741...32D} D{\'{\i}}az-Santos, T., Charmandaris, V., Armus, L., et al.\ 2011, \apj, 
741, 32 
\bibitem[D{\'{\i}}az-Santos et al.(2010)]{2010ApJ...723..993D} D{\'{\i}}az-Santos, T., Charmandaris, V., Armus, L., et al.\ 2010, \apj, 
723, 993 
\bibitem[D{\'{\i}}az-Santos et al.(2008)]{2008ApJ...685..211D} D{\'{\i}}az-Santos, T., Alonso-Herrero, A., Colina, L., et al.\ 2008, \apj, 
685, 211 
\bibitem[Draine et al.(2007)]{2007ApJ...663..866D} Draine, B.~T., et al.\ 2007, \apj, 663, 866
\bibitem[Elbaz et al.(2011)]{2011A&A...533A.119E} Elbaz, D., Dickinson, M., Hwang, H.~S., et al.\ 2011, \aap, 533, A119 
\bibitem[Eisenhauer et al.(2003)]{2003SPIE.4841.1548E} Eisenhauer, F., et  al.\ 2003, \procspie, 4841, 1548
\bibitem[Engelbracht et al.(2008)]{2008ApJ...678..804E} Engelbracht, C.~W., Rieke, G.~H., Gordon, K.~D., Smith, J.-D.~T., Werner, M.~W., Moustakas, J., Willmer, C.~N.~A., \& Vanzi, L.\ 2008, \apj, 678, 804
\bibitem[Epinat et  al.(2009)]{2009A&A...504..789E} Epinat, B., et al.\ 2009, \aap, 504, 789   
\bibitem[Erb et al.(2004)]{2004ApJ...612..122E} Erb, D.~K., Steidel, C.~C.,  Shapley, A.~E., Pettini, M., \& Adelberger, K.~L.\ 2004, \apj, 612, 122
\bibitem[Erb et al.(2006)]{2006ApJ...647..128E} Erb, D.~K., Steidel, C.~C., Shapley, A.~E., Pettini, M., Reddy, N.~A., \& Adelberger, K.~L.\ 2006 a, \apj, 647, 128 
\bibitem[Erb et al.(2006)]{2006ApJ...646..107E} Erb, D.~K., Steidel, C.~C., Shapley, A.~E., Pettini, M., Reddy, N.~A., \& Adelberger, K.~L.\ 2006 b, \apj, 646, 107 
\bibitem[Erb et al.(2006)Metal]{2006ApJ...644..813E} Erb, D.~K., Shapley, A.~E., Pettini, M., Steidel, C.~C., Reddy, N.~A., \& Adelberger, K.~L.\ 2006 c, \apj, 644, 813 
\bibitem[Farrah et al.(2008)]{2008ApJ...677..957F} Farrah, D., Lonsdale, C.~J., Weedman, D.~W., et al.\ 2008, \apj, 677, 957 
\bibitem[F{\"o}rster Schreiber et al.(2011)]{2011ApJ...731...65F}  F{\"o}rster Schreiber, N.~M., Shapley, A.~E., Erb, D.~K., Genzel, R.,  Steidel, C.~C., Bouch{\'e}, N., Cresci, G., \& Davies, R.\ 2011, \apj, 731, 65 
\bibitem[F{\"o}rster Schreiber et al.(2009)]{2009ApJ...706.1364F}  F{\"o}rster Schreiber, N.~M., et al.\ 2009, \apj, 706, 1364 
\bibitem[F{\"o}rster Schreiber et al.(2006)]{2006ApJ...645.1062F} F{\"o}rster Schreiber, N.~M., et al.\ 2006, \apj, 645, 1062 
\bibitem[Garc{\'{\i}}a-Mar{\'{\i}}n et  al.(2009)]{2009aA&A...505.1319G} Garc{\'{\i}}a-Mar{\'{\i}}n, M., Colina, L., Arribas, S., \& Monreal-Ibero, A.\ 2009a, \aap, 505, 1319 (GM09a) 
\bibitem[Garc{\'{\i}}a-Mar{\'{\i}}n et  al.(2009)]{2009bA&A...505.1017G} Garc{\'{\i}}a-Mar{\'{\i}}n, M., Colina, L., \& Arribas, S.\ 2009b, \aap, 505, 1017 (GM09b) 
\bibitem[Garc{\'{\i}}a-Mar{\'{\i}}n (2007)]{2007}  Garc{\'{\i}}a-Mar{\'{\i}}n, M.\ 2007, Ph.Thesis, Universidad Autonoma de Madrid 
\bibitem[Garc{\'{\i}}a-Mar{\'{\i}}n et al.(2006)]{2006ApJ...650..850G}  Garc{\'{\i}}a-Mar{\'{\i}}n, M., Colina, L., Arribas, S., Alonso-Herrero, A., \& Mediavilla, E.\ 2006, \apj, 650, 850 
\bibitem[2008]{Gen08} Genzel, R. et al. 2008,\apj, 687, 59
\bibitem[2010]{Gen10} Genzel, R. et al. 2010, \mnras, 407,209
\bibitem[Iono et al.(2009)]{2009ApJ...695.1537I} Iono, D., et al.\ 2009, \apj, 695, 1537
\bibitem[Jones et al.(2010)]{2010MNRAS.404.1247J} Jones, T.~A., Swinbank, A.~M., Ellis, R.~S., Richard, J., \& Stark, D.~P.\ 2010, \mnras, 404, 1247
\bibitem[Kartaltepe et al.(2011)]{2011arXiv1110.4057K} Kartaltepe, J.~S., Dickinson, M., Alexander, D.~M., et al.\ 2011, arXiv:1110.4057  
\bibitem[Kennicutt(1998)]{1998ARA&A..36..189K} Kennicutt, R.~C., Jr.\ 1998, \araa, 36, 189 
\bibitem[Kim et al.(1995)]{1995ApJS...98..129K} Kim, D.-C., Sanders, D.~B., Veilleux, S., Mazzarella, J.~M., \& Soifer, B.~T.\ 1995, \apjs, 98, 129 
\bibitem[Kim \& Sanders(1998)]{1998ApJS..119...41K} Kim, D.-C., \& Sanders, D.~B.\ 1998, \apjs, 119, 41
\bibitem[Kim et al.(2002)]{2002ApJS..143..277K} Kim, D.-C., Veilleux, S., \& Sanders, D.~B.\ 2002, \apjs, 143, 277  
\bibitem[Larkin et al.(2006)]{2006SPIE.6269E..42L} Larkin, J., et al.\  2006, \procspie, 6269,  42 
\bibitem[Law et al.(2009)]{2009ApJ...697.2057L} Law, D.~R., Steidel, C.~C., Erb, D.~K., Larkin, J.~E., Pettini, M., Shapley, A.~E.,  \& Wright, S.~A.\ 2009, \apj, 697, 2057
\bibitem[Lawrence et al.(1999)]{1999MNRAS.308..897L} Lawrence, A., Rowan-Robinson, M., Ellis, R.~S., et al.\ 1999, \mnras, 308, 897 
\bibitem[Leech et al.(1994)]{1994MNRAS.267..253L} Leech, K.~J., Rowan-Robinson, M., Lawrence, A., \& Hughes, J.~D.\ 1994, \mnras, 267, 253  
\bibitem[LeF$\grave{e}$vre et al.(2003)]{lef03} LeF$\grave{e}$vre, O., et al.2003, \procspie, 4841, 1670
\bibitem[Maiolino et al.(2008)]{2008A&A...488..463M} Maiolino, R., Nagao, T., Grazian, A., et al.\ 2008, \aap, 488, 463
\bibitem[Melnick \& Mirabel(1990)]{1990A&A...231L..19M} Melnick, J., \& Mirabel, I.~F.\ 1990, \aap, 231, L19  
\bibitem[Monreal-Ibero et al.(2010)]{2010A&A...517A..28M} Monreal-Ibero, A., Arribas, S., Colina, L., Rodr{\'{\i}}guez-Zaur{\'{\i}}n, J., Alonso-Herrero, A., \& Garc{\'{\i}}a-Mar{\'{\i}}n, M.\ 2010, \aap, 517, A28 
\bibitem[Monreal-Ibero et al.(2006)]{2006ApJ...637..138M} Monreal-Ibero, A., Arribas, S., \& Colina, L.\ 2006, \apj, 637, 138 
\bibitem[Muzzin et al.(2010)]{2010ApJ...725..742M} Muzzin, A., van Dokkum, P., Kriek, M., Labb{\'e}, I., Cury, I., Marchesini, D., \& Franx, M.\ 2010, \apj, 725, 742 
\bibitem[Nardini et al.(2010)]{2010MNRAS.403.1131N} Nardini, M., Ghisellini, G., Ghirlanda, G., \& Celotti, A.\ 2010, \mnras, 403, 1131 
\bibitem[Nelson et al.(2012)]{2012arXiv1202.1822N} Nelson, E.~J., van Dokkum, P.~G., Brammer, G., et al.\ 2012, arXiv:1202.1822 
\bibitem[Papovich et al.(2007)]{2007ApJ...668...45P} Papovich, C., et al.\ 2007, \apj, 668, 45
\bibitem[Peng et al.(2010)]{2010AJ....139.2097P} Peng, C.~Y., Ho, L.~C., Impey, C.~D., \& Rix, H.-W.\ 2010, \aj, 139, 2097 
\bibitem[2005]{Pe05} P\'erez-Gonz\'alez, P.G. et al. 2005,\apj, 630, 82
\bibitem[Pettini \& Pagel(2004)]{2004MNRAS.348L..59P} Pettini, M., \& Pagel, B.~E.~J.\ 2004, \mnras, 348, L59 
\bibitem[Pope et al.(2006)]{2006MNRAS.370.1185P} Pope, A., et al.\ 2006, \mnras, 370, 1185 
\bibitem[Rigby et al.(2008)]{2008ApJ...675..262R} Rigby, J.~R., et al.\ 2008, \apj, 675, 262
\bibitem[Rodighiero2011)]{2011astro-ph}Rodighiero, G. et al.\ 2011, astro-ph 2011arXiv1108.0933R
\bibitem[Rodr{\'{\i}}guez-Zaur{\'{\i}}n et  al.(2011)]{2011A&A...527A..60R} Rodr{\'{\i}}guez-Zaur{\'{\i}}n, J., Arribas, S., Monreal-Ibero, A., Colina, L., Alonso-Herrero, A., \& Alfonso-Garz{\'o}n, J.\ 2011, \aap, 527, A60 (RZ11)
\bibitem[Rosales-Ortega et  al.(2012)]{2012A&A...539A..73R} Rosales-Ortega, F.~F., Arribas, S., \& Colina, L.\ 2012, \aap, 539, A73 
\bibitem[Rujopakarn et al.(2011)]{2011ApJ...726...93R} Rujopakarn, W., Rieke, G.~H., Eisenstein, D.~J., \& Juneau, S.\ 2011, \apj, 726, 93 
\bibitem[Rupke et al.(2008)]{2008ApJ...674..172R} Rupke, D.~S.~N., Veilleux, S., \& Baker, A.~J.\ 2008, \apj, 674, 172
\bibitem[Sanders et al.(1988)]{1988ApJ...328L..35S} Sanders, D.~B., Soifer, B.~T., Elias, J.~H., Neugebauer, G., \& Matthews, K.\ 1988, \apjl, 328, L35 
\bibitem[Sanders et al.(2003)]{2003AJ....126.1607S} Sanders, D.~B., Mazzarella, J.~M., Kim, D.-C., Surace, J.~A., 
\& Soifer, B.~T.\ 2003, \aj, 126, 1607 
\bibitem[Scott et al.(2009)]{2009ApJ...691L.119S} Scott, P., Asplund, M., Grevesse, N., \& Sauval, A.~J.\ 2009, \apjl, 691, L119  
\bibitem[Shapiro et al.(2009)]{2009ApJ...701..955S} Shapiro, K.~L., et al.\  2009, \apj, 701, 955 
\bibitem[Smail et al.(1997)]{1997ApJ...490L...5S} Smail, I., Ivison, R.~J., \& Blain, A.~W.\ 1997, \apjl, 490, L5
\bibitem[Sobral et al.(2012)]{2012arXiv1202.3436S} Sobral, D., Smail, I., Best, P.~N., et al.\ 2012, arXiv:1202.3436 
\bibitem[Steidel et al.(2004)]{2004ApJ...604..534S} Steidel, C.~C., Shapley, A.~E., Pettini, M., Adelberger, K.~L., Erb, D.~K., Reddy, N.~A., \& Hunt, M.~P.\ 2004, \apj, 604, 534
\bibitem[Swinbank et al.(2004)]{2004ApJ...617...64S} Swinbank, A.~M., Smail, I., Chapman, S.~C., et al.\ 2004, \apj, 617, 64 
\bibitem[Swinbank et al.(2006)]{2006MNRAS.371..465S} Swinbank, A.~M., Chapman, S.~C., Smail, I., et al.\ 2006, \mnras, 371, 465
\bibitem[Tacconi et al.(2006)]{2006ApJ...640..228T} Tacconi, L.~J., Neri, R., Chapman, S.~C., et al.\ 2006, \apj, 640, 228 
\bibitem[Tacconi et al.(2008)]{2008ApJ...680..246T} Tacconi, L.~J., Genzel, R., Smail, I., et al.\ 2008, \apj, 680, 246 
\bibitem[Takagi et  al.(2010)]{2010A&A...514A...5T} Takagi, T., et al.\ 2010, \aap, 514, A5 
\bibitem[Tecza et al.(2004)]{2004ApJ...605L.109T} Tecza, M., Baker, A.~J., Davies, R.~I., et al.\ 2004, \apjl, 605, L109
\bibitem[Veilleux et al.(1995)]{1995ApJS...98..171V} Veilleux, S., Kim, D.-C., Sanders, D.~B., Mazzarella, J.~M., \& Soifer, B.~T.\ 1995, \apjs, 98, 171 
\bibitem[Veilleux et al.(1999)]{1999ApJ...522..113V} Veilleux, S., Kim, D.-C., \& Sanders, D.~B.\ 1999, \apj, 522, 113 
\bibitem[Veilleux et al.(2002)]{2002ApJS..143..315V} Veilleux, S., Kim, D.-C., \& Sanders, D.~B.\ 2002, \apjs, 143, 315 
\bibitem[Wisnioski et al.(2011)]{2011MNRAS.417.2601W} Wisnioski, E., Glazebrook, K., Blake, C., et al.\ 2011, \mnras, 417, 2601 
\bibitem[Wright et al.(2009)]{2009ApJ...699..421W} Wright, S.~A., Larkin, J.~E., Law, D.~R., Steidel, C.~C., Shapley, A.~E., \& Erb, D.~K.\ 2009, \apj, 699, 421 

\end{thebibliography}
\end{document}